\documentclass[final]{jfm}

\usepackage{graphicx}
\usepackage{natbib}

\ifCUPmtlplainloaded \else
  \checkfont{eurm10}
  \iffontfound
    \IfFileExists{upmath.sty}
      {\typeout{^^JFound AMS Euler Roman fonts on the system,
                   using the 'upmath' package.^^J}%
       \usepackage{upmath}}
      {\typeout{^^JFound AMS Euler Roman fonts on the system, but you
                   dont seem to have the}%
       \typeout{'upmath' package installed. JFM.cls can take advantage
                 of these fonts,^^Jif you use 'upmath' package.^^J}%
       \providecommand\upi{\pi}%
      }
  \else
    \providecommand\upi{\pi}%
  \fi
\fi

\ifCUPmtlplainloaded \else
  \checkfont{msam10}
  \iffontfound
    \IfFileExists{amssymb.sty}
      {\typeout{^^JFound AMS Symbol fonts on the system, using the
                'amssymb' package.^^J}%
       \usepackage{amssymb}%
         
         \let\geq=\geqslant
      }{}
  \fi
\fi

\ifCUPmtlplainloaded \else
  \IfFileExists{amsbsy.sty}
    {\typeout{^^JFound the 'amsbsy' package on the system, using it.^^J}%
     \usepackage{amsbsy}}
    {\providecommand\boldsymbol[1]{\mbox{\boldmath $##1$}}}
\fi

\providecommand\bnabla{\boldsymbol{\nabla}}
\providecommand\bcdot{\boldsymbol{\cdot}}

\newsavebox{\astrutbox}
\sbox{\astrutbox}{\rule[-5pt]{0pt}{20pt}}

\newcommand\p{\ensuremath{\partial}}
\newcommand\D{\ensuremath{\mathrm{D}}}
\newcommand\etal{\mbox{\textit{et al.}}}

\newcommand\eg{e.g.}
\newcommand\be{\mathbf{e}}
\newcommand\floq{f}
\newcommand\grav{\mathbf{G}}
\newcommand\bk{\mathbf{k}}

\newcommand\eikx{\mathrm{e}^{\mathrm{i}{\bf k}{\cdot}{\bf x}}}

\newcommand\kgpermc{\nobreak\mbox{$\;$kg\,m$^{-3}$}}
\newcommand\Patimess{\nobreak\mbox{$\;$Pa\,s}}
\newcommand\mPatimess{\nobreak\mbox{$\;$mPa\,s}}
\newcommand\Nperm{\nobreak\mbox{$\;$N\,m$^{-1}$}}
\newcommand\mNperm{\nobreak\mbox{$\;$mN\,m$^{-1}$}}
\newcommand\mperssq{\nobreak\mbox{$\;$m\,s$^{-2}$}}
\newcommand\Hz{\nobreak\mbox{$\;$Hz}}
\newcommand\mm{\nobreak\mbox{$\;$mm}}
\newcommand\cm{\nobreak\mbox{$\;$cm}}
\newcommand\permm{\nobreak\mbox{$\;$mm$^{-1}$}}
\newcommand\s{\nobreak\mbox{$\;$s}}

\title[Numerical simulation of Faraday waves]{Numerical simulation of Faraday waves}

\author[N. P\'erinet, D. Juric and L. S. Tuckerman]%
{N\ls I\ls C\ls O\ls L\ls A\ls S\ns  P\ls \'E\ls R\ls I\ls N\ls E\ls T$^1$\ls ,\ns
D\ls A\ls M\ls I\ls R\ns  J\ls U\ls R\ls I\ls C$^2$
\and L\ls A\ls U\ls R\ls E\ls T\ls T\ls E\ns S\ls .\ns T\ls U\ls C\ls K\ls E\ls R\ls M\ls A\ls N$^1$\break}

\affiliation{$^1$Laboratoire de Physique et M\'ecanique des Milieux H\'et\'erog\`enes (PMMH),\\
Ecole Sup\'erieure de Physique et de 
Chimie Industrielles de la Ville de Paris (ESPCI),\\
Centre National de la Recherche Scientifique (CNRS), UMR 7636,\\
Universit\'e Paris 6 et Paris 7,\\
10 rue Vauquelin, 75231 Paris Cedex 5, France
\\[\affilskip]
$^2$Laboratoire d'Informatique pour la M$\acute{\textmd{e}}$canique et les Sciences de l'Ing\'enieur (LIMSI),\\
Centre National de la Recherche Scientifique (CNRS), UPR 3251, \\
BP133, 91403 Orsay Cedex, France}

\begin{document}

\maketitle

\begin{abstract}
We simulate numerically the full dynamics of Faraday waves in three dimensions
for two incompressible and immiscible viscous fluids. The Navier--Stokes
equations are solved using a finite-difference projection method coupled with
a front-tracking method for the interface between the two fluids. 
The critical accelerations
and wavenumbers, as well as the temporal behaviour at onset are compared with
the results of the linear Floquet analysis of Kumar \& Tuckerman~(\textit{J.~Fluid Mech.}, vol. 279, 1994, p. 49).
The finite-amplitude results are compared with the experiments of Kityk \etal~(\textit{Phys.~Rev.}~E, vol. 72, 2005, p. 036209).
In particular, we reproduce the detailed spatio-temporal spectrum of both square and hexagonal patterns within experimental uncertainty.
We present the first calculations of a three-dimensional velocity field arising from the Faraday instability for a hexagonal pattern 
as it varies over its oscillation period.

\end{abstract}

\maketitle

\section{Historical introduction} 

The Faraday experiment consists of shaking
vertically a container holding two immiscible fluids (the lighter of which
can be air) thereby inducing oscillations of the fluids and the interface
between them. Beyond a certain threshold, the interface can form many kinds of
standing wave patterns, including crystalline patterns and others which are more
complex. This phenomenon was first studied by \cite{Far1831}
who noticed that the vibration frequency of the interface was half that of the
forcing. The results of Faraday were confirmed by 
Rayleigh \cite{Ray1883a,Ray1883b}. 
\cite{BU1954} carried out the first theoretical linear analysis of the Faraday 
waves, restricted to inviscid fluids. They decomposed the fluid motion into normal modes of the container and 
showed that the evolution equation of each mode reduced to a Mathieu equation 
whose stability diagram is well known. 

In the 1990s, new behaviours of the interface were
discovered, such as quasi-crystalline eight-fold patterns seen by
\cite{CAL1992}. By introducing a forcing which is the sum of two periodic
functions with commensurable frequencies, \cite{EF1994} were able to produce
twelve-fold quasi-patterns. Triangular patterns were observed by \cite{M1993}
and superlattice patterns by \cite{KPG1998}, also using two-frequency forcing.
Spatio-temporal chaos was studied by \cite{KG1996}, who also surveyed the
occurrence of lattice patterns -- stripes, squares or hexagons -- as a
function of viscosity and frequency.  \cite{BWW1997} 
demonstrated the dependence of the pattern on the depth of the layer.
In addition to patterns or
quasi-patterns, very localized circular waves called oscillons may occur,
as seen by \cite{LAF1996}.  The Faraday instability is the first macroscopic
system in which such structures have been observed. These discoveries endow
the Faraday instability with a very great fundamental interest for
understanding the natural formation of patterns.

A number of theoretical or semi-numerical analyses were inspired by these
experiments. \cite{KT1994} extended the linear stability analysis of \cite{BU1954} to viscous fluids.  This analysis was experimentally confirmed by
\cite{BEMJ1995} and used by \cite{K1996} to predict cases in which the
response would be harmonic rather than subharmonic.  The method was extended by
\cite{BET1996} to calculate the stability tongues in the case of two-frequency
forcing.  Integral equation formulations of the viscous linear stability 
problem were derived by \cite{BF1995} and \cite{MWWAK1997}, who also 
studied the harmonic response case.  
\cite{CT98} used the lubrication approximation and the WKB method to 
study shallow viscous layers, obtaining a Mathieu equation that was 
later used by \cite{HDUS2006} to derive analytic results about the response to 
multifrequency forcing.

Linear analysis provides no information about the shape of the patterns which
appear; other means are necessary to understand the occurence of a given
pattern or the amplitude of stabilization.  Weakly nonlinear approximations
have been derived from the Navier--Stokes equations by Vi\~{n}als and
co-workers, \eg~\cite{ZV1997}, \cite{CV1999} and by \cite{SG2007},
focusing on the competition between different patterns.  \cite{VKM2001}
derived equations governing the interaction between Faraday waves and the mean
flow.  There has been a great deal of analysis of lattices, superlattices and
quasi-patterns using equivariant dynamical systems theory, as well as model
equations designed to produce specific patterns, e.g.  \cite{PTS2004}.  The
approximation of quasipatterns in spatially periodic domains has also been
addressed in \cite{RS09}.

Investigation of the full nonlinear viscous problem, however, requires
numerical simulations, of which there have been very few up to now,
specifically those of \cite{CW2000}, \cite{C2002}, \cite{MC2001}, \cite{VLK2002},
\cite{UGS2003} and \cite{O'Co2008}.  With the exception of \cite{O'Co2008},
all previous simulations have been two-dimensional.  The most extensive
simulation thus far has been that of \cite{CW2000} and \cite{C2002}, who used
a finite-difference method applied to a boundary-fitted time-dependent
coordinate system.  At each timestep, the surface is advected and a new 2D
grid, adapted to the surface, is recomputed.  The amplitude of their
numerically computed Faraday waves confirmed the weakly nonlinear analysis of
\cite{CV1999}, including their prediction of a range of subcriticality.  Their
calculations also predicted qualitatively new phenomena, such as disconnected
solution branches and slow modulated dynamics.

\cite{MC2001} used a method similar to that of \cite{CW2000} and \cite{C2002}.
Although they reproduced some features of the experiments by \cite{LAF1996},
their calculations were limited to accelerations only 0.5 \%
above critical.  The investigation by Ubal \etal~(2003) focused on the
influence of liquid depth in two-dimensional simulations using a Galerkin
finite-element method in transformed coordinates.  In addition to comparing
their linear stability predictions with those of \cite{BU1954} and \cite{KT1994},
they calculated instantaneous surface profiles and velocity
fields, as well as the temporal evolution and spectrum. Valha \etal~(2002)
examined the response of a liquid layer in a vertical cylindrical vessel using
the MAC (Marker-and-Cell) method of \cite{HW1965}.  Surface tension was
treated by the continuum surface force model of
\cite{BKZ1992}. \cite{O'Co2008} conducted numerical simulations using an ALE (Arbitrary Lagrangian-Eulerian) spectral-element code in both two and three dimensions; a visualization of a
square pattern was presented.

The hexagonal patterns, quasi-patterns and oscillons which motivate our
investigation are intrinsically three-dimensional, and have never 
been calculated numerically from the fluid-dynamical equations.
Here we report on the results of fully nonlinear,
three-dimensional simulations of Faraday waves using a finite-difference
front-tracking method.  
In the classic Faraday problem the lighter fluid is usually taken to be air
whose effects can be neglected.  However, in contrast to the previously cited
investigations, the numerical method described here solves the Navier--Stokes
equations for the general case of two distinct superposed fluids.  The
capability of the method to simulate the motion of both fluids is important in
that it permits comparison of numerical results with those of certain
experimental configurations, namely those of \cite{KEMW2005} where the lighter
fluid cannot be ignored.
These experiments were the first to
provide quantitative measurements of the complete spatio-temporal Fourier
spectrum of Faraday waves and thus form an excellent basis for quantitative
comparison with our numerical results in the nonlinear finite amplitude regime, i.e.
interfaces with steep slopes.

In the next two sections of this article, we present the hydrodynamic
equations that govern the Faraday instability and then describe the
computational method. The two sections following these are dedicated to the
comparison of our results with the linear theory of \cite{KT1994} and with the
experiments of Kityk \etal~(2005, 2009). After comparing numerical and experimental
spatio-temporal spectra for squares and hexagons, we present the
three-dimensional velocity field for the hexagonal pattern.

\section{Equations of motion}

The mathematical model of the Faraday experiment consists of two incompressible and immiscible viscous fluids in a three-dimensional domain
$\textbf{x}=(x,y,z)\in \Re^2\times[0,h]$, bounded at $z=0$ and $z=h$ by flat
walls. The two fluids, each uniform and of densities $\rho_1$, $\rho_2$ and viscosities
$\mu_1$ and $\mu_2$,  initially form two superposed horizontal layers with an interface between them.
This two-dimensional interface is defined by  $\textbf{x}'=\left(x,y,\zeta(x,y,t)\right)$. 
Within the parameter range we wish to simulate, the height $\zeta$ remains a single-valued function of $(x,y,t)$.

The container is shaken vertically in $z$. In the reference frame of the container the boundary conditions for the fluid velocities $\textbf{u}=(u,v,w)$ are
\begin{subequations}\label{eq2-1}
\begin{eqnarray}
\textbf{u}(x,y,0,t)&=&0, \label{eq2-1a}\\
\textbf{u}(x,y,h,t)&=&0. \label{eq2-1b}
\end{eqnarray}
\end{subequations}
The gravitational acceleration $g$ is augmented by a temporally periodic inertial acceleration
\begin{equation}\label{eq2-2}
\grav=(a\cos(\omega t)-g)\textbf{e}_z,
\end{equation}
where $a$ is the amplitude of the forcing and $\omega$ is its frequency.

The Navier--Stokes equations for incompressible, Newtonian fluids are
\begin{subequations}\label{eq2-3}
\begin{equation}
\displaystyle{\rho\frac{\D \textbf{u}}{\D t}=-\bnabla p+\rho\grav}
\displaystyle{+\bnabla\bcdot\mu\left(\bnabla\textbf{u}+\bnabla\textbf{u}^{\rm
    T}\right)+\textbf{s},} \label{eq2-3b} 
\end{equation}\begin{equation}
\bnabla\bcdot\textbf{u}=0. \label{eq2-3a}
\end{equation}
\end{subequations}
Here $p$ is the pressure and $\textbf{s}$ is the capillary force (per unit volume) and is defined below.
Equations (\ref{eq2-3b}) and (\ref{eq2-3a}) are valid for the entire domain, including the interface, in spite of the fact that the density and viscosity change discontinuously and the surface tension acts only at the interface.  In this single-fluid formulation, the density and viscosity fields are defined in terms of the densities and viscosities of the two fluids
\begin{subequations}\label{eq2-4}
\begin{eqnarray}
\rho&=&\rho_{1}+(\rho_{2}-\rho_{1})H, \label{eq2-4a} \\
\mu&=&\mu_{1}+(\mu_{2}-\mu_{1})H, \label{eq2-4b}
\end{eqnarray}
\end{subequations}
with the aid of a Heaviside function,
\begin{equation}\label{eq2-7}
H\left(\textbf{x}-\textbf{x}'\right)=\left\{
\begin{array}{ll}
0 \textmd{ if }z<\zeta(x,y,t)
\\
1 \textmd{ if }z\geq \zeta(x,y,t)
\end{array},
\right.
\end{equation}
where we recall that $\textbf{x}=(x,y,z)$ is a point anywhere in the three-dimensional volume and $\textbf{x}'=\left(x,y,\zeta(x,y,t)\right)$ is the vertical projection of $\textbf{x}$ onto the interface. The capillary force is
\begin{equation}\label{eq2-8}
\textbf{s}=\int_{S'(t)}{\sigma\kappa\,\textbf{n}\,\delta\left(\textbf{x}-\textbf{x}'\right)\mathrm{d}S,}
\end{equation}
where $\sigma$ is the surface tension coefficient, assumed to be constant, $\textbf{n}$ is the unit normal to the interface (directed into the upper fluid) and $\kappa$ its curvature. $\delta\left(\textbf{x}-\textbf{x}'\right)$ is a three-dimensional Dirac distribution that is nonzero only where $\textbf{x}=\textbf{x}'$.  $S'(t)$ is the surface defined by the instantaneous position of the interface.

To complete the system of equations we need an expression for the motion of the interface. One such expression can be easily derived by noting that mass conservation in an incompressible flow requires $\D\rho/\D t=0$, which in view of the discontinuous density field (\ref{eq2-4a}), is equivalent to
\begin{equation}\label{eq2-9}
\D H/\D t=0.
\end{equation}
Thus the interface is represented implicitly by $H$ and advected by material motion of the fluid.

\section{Computational methods}

The computational domain is a rectangular parallelepiped, horizontally
periodic in $x$ and $y$ and bounded in $z$ by flat walls for which we impose
no-slip boundary conditions.  The entire domain is discretized by a uniform
fixed three-dimensional finite-difference mesh. This mesh has a standard
staggered MAC cell arrangement \cite[][]{HW1965} where the $u$, $v$ and $w$
velocity nodes are located on the corresponding cell faces and scalar
variables are located at the cell centres.  Each cell is of dimension $\Delta
x \times \Delta y \times \Delta z$.

\begin{figure}
\begin{center}
\includegraphics[width=12cm]{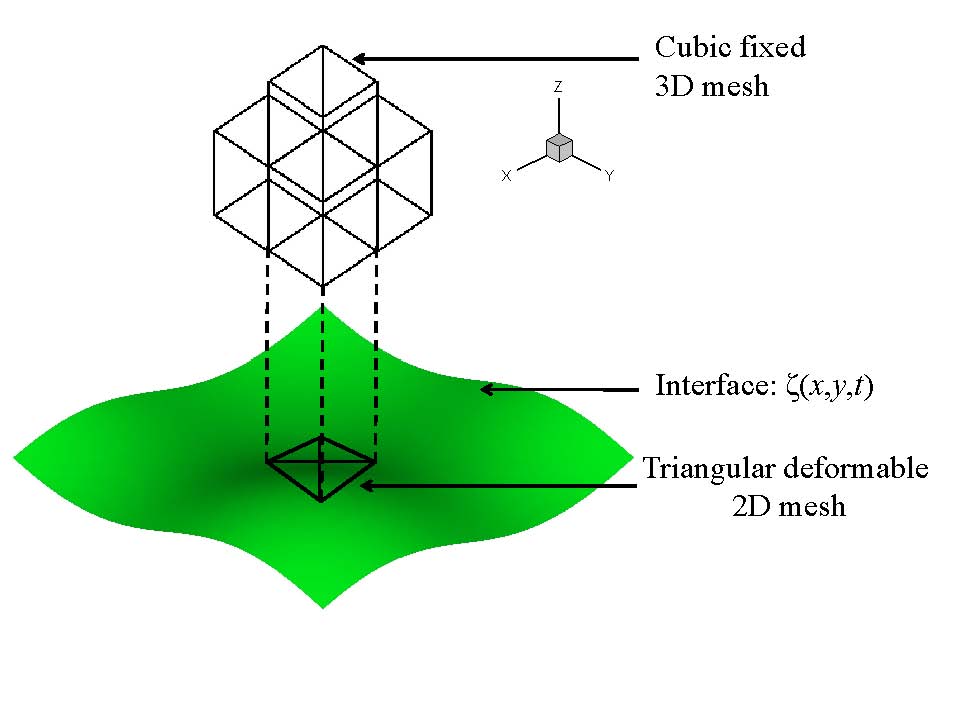}
\caption{Spatial discretization of the domain.}
\label{figdomain}
\end{center}
\end{figure}

Within the domain, the two distinct immiscible fluids are separated by a
two-dimensional interface which is discretized by a second mesh as sketched in
figure \ref{figdomain}.  This moving and deformable mesh is composed of
triangular elements whose motion is treated by a
front-tracking/immersed-boundary method \cite[][]{Pes1977, T&al2001}.
Because we have assumed that $\zeta(x,y,t)$ is single valued, the nodes of the
mesh can be fixed in $x$ and $y$ and only their vertical displacements need to
be calculated, which is a considerable simplification to the general
front-tracking method.

After setting appropriate initial and boundary conditions, the computational
solution algorithm for each timestep is composed of three main
phases. First, the interface is advected and the density and viscosity fields
updated according to the new interface position. The capillary force
$\textbf{s}$ is then calculated. Finally, the velocity and pressure are found
by means of a standard projection method. Each of these steps is described
below.

\subsection{Advection of the interface}

Purely Eulerian interface methods such as Volume of Fluid (Hirt \& Nichols 1981) or Level Set (Osher \& Sethian 1988) use a form of (\ref{eq2-9}) to advect a scalar field such as $H$, or a level-set function that implicitly represents the interface. However, in the front-tracking approach that we use here, the interface markers themselves (the nodes of the triangular mesh) are advected.  $H$ is then constructed from the position and geometry of the interface.  Taking (\ref{eq2-9}) as a starting point, we develop an equivalent expression for the vertical displacement of the triangular interface mesh.

The material derivative of (\ref{eq2-7}) gives:
\begin{equation}\label{eq3-9}
\frac{\D H}{\D t}=\bnabla H\bcdot \frac{\D \textbf{x}}{\D t} +\bnabla' H\bcdot \frac{\D \textbf{x}'}{\D t},
\end{equation}
where $\displaystyle{\bnabla'=\p_{\textbf{x}'}}$ and 
\begin{equation}\label{gradh}
\bnabla H=-\bnabla'H=\int_{S'(t)} \textbf{n}\delta(\textbf{x}-\textbf{x}')\mathrm{d}S.
\end{equation}
(For a derivation of (\ref{gradh}) see Tryggvason \etal~2001.)  Factoring (\ref{eq3-9}) by $\bnabla H$
\begin{equation}\label{eq3-10}
\frac{\D H}{\D t}=\bnabla H\bcdot\left(\textbf{u}-\frac{\D \textbf{x}'}{\D t}\right).
\end{equation}
The right-hand side of (\ref{eq3-10}) can only be zero everywhere, including on the surface, $\displaystyle{\textbf{x}=\textbf{x}'}$, if
\begin{equation}\label{eq3-11}
\frac{\D \textbf{x}'}{\D  t}=\textbf{u}(\textbf{x}',t),
\end{equation}
which establishes the material motion of the explicit interface representation $\textbf{x}'$.
Furthermore
\begin{equation}\label{eq3-12}
\frac{\p\textbf{x}'}{\p t}=\left(\textbf{I}-\bnabla\textbf{x}'\right)\bcdot\textbf{u}(\textbf{x}',t),
\end{equation}
where $\textbf{I}$ is the identity tensor. 
The specific choice of $\textbf{x}'=(x,y,\zeta(x,y,t))$ made here gives for $\bnabla\textbf{x}'$
\begin{equation}\label{eq3-13}
\left[
\begin{array}{ccc}
1 & 0 & 0
\vspace{2.5pt}
\\
0 & 1 & 0
\vspace{2.5pt}
\\
\displaystyle{\frac{\p\zeta}{\p x}} & \displaystyle{\frac{\p\zeta}{\p y}} & 0

\end{array}
\right].
\end{equation}
With this, (\ref{eq3-12}) leads to the specific displacement relations:
\begin{subequations}\label{eq3-14}
\begin{equation}\label{eq3-14a}
\displaystyle{\p x/\p t=0,}
\end{equation}
\begin{equation}\label{eq3-14b}
\displaystyle{\p y/\p t=0,}
\end{equation}
\begin{equation}\label{eq3-14c}
\displaystyle{\frac{\p\zeta}{\p t}=w-u\frac{\p\zeta}{\p x}-v\frac{\p\zeta}{\p y}.}
\end{equation}
\end{subequations}
The application of this advection to the triangular mesh we use for tracking the interface is straightforward.  At each vertex the horizontal displacement is zero and for the vertical displacement we compute a first-order approximation to (\ref{eq3-14c}):  
\begin{equation}\label{eq3-16}
\displaystyle{\frac{\zeta^{n+1}-\zeta^{n}}{\Delta t}=w^{n}({\textbf{x}'_e})-\frac{\p \zeta^{n}}{\p x}u^{n}(\textbf{x}'_e)-\frac{\p \zeta^{n}}{\p y}v^{n}(\textbf{x}'_e).}
\end{equation}
The superscripts $n$ and $n+1$ denote, respectively, the old and new time levels.  The derivatives on the right-hand side are evaluated using a simple upwind scheme which requires the usual CFL (Courant-Friedrichs-Lewy) time step restriction. The vertical displacement of the interface mesh requires knowledge of the velocities at the element nodes $\textbf{x}'_{e}$. These in general do not coincide with the Eulerian grid nodes $\textbf{x}_{ijk}$, whose indices correspond to discretized coordinates along the respective directions $x$, $y$ and $z$.  The problem of communicating Eulerian grid velocities to the element nodes is overcome by interpolation between the two grids as is typically done in front-tracking and immersed-boundary methods.  Here we use the particular interpolation
\begin{equation}\label{eq3-int}
\textbf{u}(\textbf{x}'_{e})=\sum_{ijk}{\textbf{u}(\textbf{x}_{ijk}) \delta_h(\textbf{x}_{ijk}-\textbf{x}'_{e})\Delta x \Delta y \Delta z .}
\end{equation}
The kernel $\delta_h$ is a smoothed version of the three-dimensional Dirac
delta function with compact support of four grid nodes in each direction (for
details of the front-tracking method, see Tryggvason \etal~2001, and for the 
immersed-boundary method, see Peskin 1977). In (\ref{eq3-int}) the weighted information collected from nearby Eulerian grid nodes is interpolated to a given element node.

\begin{figure}
\begin{center}
\includegraphics[width=7.8cm]{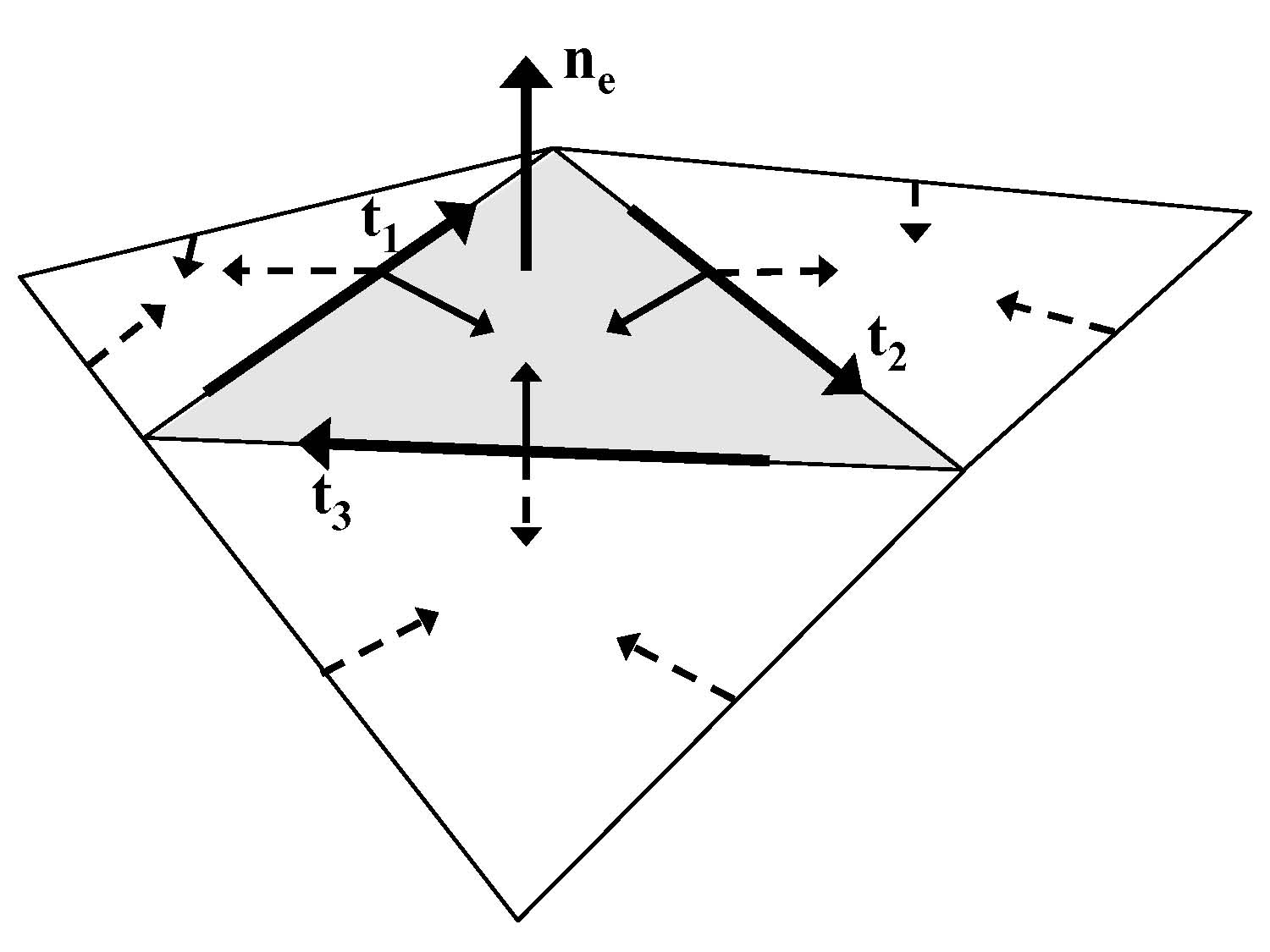}
\caption{A triangular element of the interface mesh illustrating the action of
  the capillary forces according to (\ref{eq3-1}).  For the shaded
  triangle the forces act perpendicular to the triangle's edges (unlabelled
  solid arrows), the dashed arrows are the corresponding forces on the edges
  of neighboring triangles.  The net capillary force at the shared edge
  between any two triangles (the sum of the solid and dashed vectors) is
  directed into the fluid on the concave side of the interface.  }
\label{fig0}
\end{center}
\end{figure}

We now seek to update the density and viscosity fields needed in (\ref{eq2-4}), which require $H$.
The equation for $H$, based on the updated values of $\textbf{x}'$ and 
$\textbf{n}$, is formulated by taking the divergence of (\ref{gradh}):
\begin{subequations}\label{eq3-17}
\begin{equation}\label{eq3-17a}
\displaystyle{\nabla^2 H=\bnabla\bcdot\int_{S'(t)}{\textbf{n}\,\delta\left(\textbf{x}-\textbf{x}'\right)\mathrm{d}S,}}
\end{equation}
\begin{equation}\label{eq3-17b}
H(x,y,0,t)=0,
\end{equation}
\begin{equation}\label{eq3-17c}
H(x,y,h,t)=1.
\end{equation}
\end{subequations}
The discretized version of the Poisson problem (\ref{eq3-17a}) is
\begin{equation}\label{eq3-18}
\nabla^2 H_{ijk}=\bnabla\bcdot\sum_{e}{\textbf{n}_e \delta_h(\textbf{x}_{ijk}-\textbf{x}'_{e})\Delta S_e,}
\end{equation}
where standard central differencing is used for the gradient and divergence
operators.  
This numerically calculated Heaviside function is a smoothed transition
from 0 to 1 across a distance of 4 grid cells in the direction normal to the
interface.
In contrast to (\ref{eq3-int}), the summation above serves to distribute weighted information from an element node to nearby Eulerian grid nodes.  Since an element is triangular, its vertices lie in the same plane, its normal vector is unique and the three tangent vectors are simple to calculate:
\begin{equation}\label{eq3-19}
\textbf{n}_e=\frac{\textbf{t}_2\times\textbf{t}_1}{||\textbf{t}_2\times\textbf{t}_1||}
\end{equation}
where $\textbf{t}_1$ and $\textbf{t}_2$ are the tangents on two distinct edges of the triangle (see the sketch in figure \ref{fig0}).  We solve (\ref{eq3-18}) by fast Fourier transform. Finally, $\rho^{n+1}$ and $\mu^{n+1}$ are updated using (\ref{eq2-4a}) and (\ref{eq2-4b}).

\subsection{Capillary force}

From (\ref{eq2-8}), the capillary force involves the curvature of the interface and its normal vector. However, from a computational point of view, curvature is a difficult quantity to compute accurately.  It is more accurate and physically appealing to calculate the force pulling on the edge of each individual triangular surface element and then sum the contributions for all the elements over the surface.  For a given surface element $e$ of surface area $\delta A$ and perimeter $\delta l$, we can write:
\begin{eqnarray}
\textbf{s}_e&=&\sigma\int_{\delta A}\kappa\textbf{n}\ \mathrm{d}A, \nonumber\\
&=&\sigma\int_{\delta A}(\textbf{n}\times\bnabla)\times\textbf{n}\ \mathrm{d}A,
\label{eq3-1}\\
&=&\sigma\oint_{\delta l}\textbf{t}\times\textbf{n}\ \mathrm{d}l,\nonumber
\end{eqnarray}
where the last integral represents the sum of the capillary forces exerted around the element perimeter.  As sketched in figure \ref{fig0}, the directions of these forces are oriented along the surface and normal to the element's edges.  The net force at the shared edge between any two triangles (the sum of the solid and dashed vectors) is directed into the fluid on the concave side of the interface.
Following Peskin's immersed boundary method \cite[][]{Pes1977}, the discrete version of (\ref{eq2-8})  becomes
\begin{equation}\label{eq3-2}
\textbf{s}_{ijk}=\sum_{e}{\textbf{s}_e \delta_h(\textbf{x}_{ijk}-\textbf{x}'_{e})},
\end{equation}
where we use the same smoothed $\delta_h$ as in (\ref{eq3-int}) and (\ref{eq3-18}).
Thereby, several interfacial elements contribute to the calculation of the force applied to a single Eulerian node, and a single element influences more than one Eulerian node.  

\subsection{Solution of the Navier--Stokes equations}

The Navier--Stokes equations are solved by a projection method
(Chorin 1968; Temam 1968) with incremental pressure correction (Goda 1979) applied
to a finite-difference scheme which is first order in time and second order in space. In addition a semi-implicit scheme is chosen for the velocities to relax the stability restriction on the time step due to viscous diffusion. 
All spatial derivative operators are evaluated using standard centred differences, except in the nonlinear term where we use a second-order ENO (Essentially-Non-Oscillatory) scheme \cite[][]{SO1989, S&al1998}.  (For an overview of projection methods for the incompressible Navier--Stokes equations, see Guermond, Minev \& Shen 2006.) The time stepping algorithm is thus
\begin{equation}\label{eq3-3}
\begin{array}{l}
\displaystyle{\frac{\textbf{u}^{n+1}-\textbf{u}^{n}}{\Delta t}=-\textbf{u}^{n}\bcdot\bnabla \textbf{u}^{n}+\frac{1}{\rho^{n+1}}\bnabla\bcdot\mu^{n+1}\left(\bnabla \textbf{u}+\bnabla \textbf{u}^{\rm{T}}\right)^{n+1}}
\vspace{3pt}
\\
\hspace{33pt}\displaystyle{-\frac{1}{\rho^{n+1}}\bnabla p^{n+1}+\frac{\textbf{s}^{n+1}}{\rho^{n+1}}+\grav^{~n+1},}
\end{array}
\end{equation}
with the boundary conditions on the top and bottom walls
\begin{equation}\label{eq3-4}
\left.\textbf{u}^{n+1}\right|_{\Gamma}=0.
\end{equation}
In (\ref{eq3-3}), $\rho$, $\mu$ and $\textbf{s}$ depend on $\textbf{x}$ via (\ref{eq2-4}--\ref{eq2-8}) and have already been updated by (\ref{eq3-16}).
We decompose the solution of (\ref{eq3-3}) in three steps. The first step is a semi-implicit calculation of an intermediate unprojected velocity $\tilde\textbf{u}$, involving only velocities and their gradients:
\begin{subequations}\label{eq3-5}
\begin{equation}\label{eq3-5a}
\displaystyle{\frac{\tilde\textbf{u}-\textbf{u}^{n}}{\Delta t}=-\textbf{u}^{n}\bcdot\bnabla \textbf{u}^{n}}
\displaystyle{+\frac{1}{\rho^{n+1}}\bnabla\bcdot\mu^{n+1}\left(\bnabla
  \tilde\textbf{u}+\bnabla \tilde\textbf{u}^{\rm T}\right)}\hspace{10pt}
\end{equation}
\begin{equation}\label{eq3-5b}
\left.\tilde\textbf{u}\right|_{\Gamma}=0
\end{equation}
\end{subequations}
In the second step, we include the capillary, acceleration and
old pressure gradient terms to calculate the unprojected velocity $\textbf{u}^{*}$:
\begin{equation}\label{eq3-6}
\displaystyle{\frac{\textbf{u}^*-\tilde\textbf{u}}{\Delta t}=\grav^{n+1}+\frac{\textbf{s}^{n+1}}{\rho^{n+1}}-\frac{1}{\rho^{n+1}}\bnabla p^{n}.}
\end{equation}
Finally we perform a projection step to find the divergence free velocity $\textbf{u}^{n+1}$:
\begin{subequations}\label{eq3-7}
\begin{equation}\label{eq3-7a}
\displaystyle{\frac{\textbf{u}^{n+1}-\textbf{u}^{*}}{\Delta t}=-\frac{1}{\rho^{n+1}}\bnabla \left(p^{n+1}-p^{n}\right),}
\end{equation}
\begin{equation}\label{eq3-7b}
\bnabla\bcdot\textbf{u}^{n+1}=0,
\end{equation}
\begin{equation}\label{eq3-7c}
\left.\textbf{u}^{n+1}\bcdot\textbf{n}\right|_{\Gamma}=0.
\end{equation}
\end{subequations}
Equations (\ref{eq3-7}) imply the following elliptic problem for the pressure increment
\begin{subequations}\label{eq3-8}
\begin{equation}\label{eq3-8a}
\displaystyle{\frac{\bnabla\bcdot\textbf{u}^{*}}{\Delta t}=\bnabla\bcdot\frac{1}{\rho^{n+1}}\bnabla \left(p^{n+1}- p^{n}\right),}
\hspace{34pt}
\end{equation}
\begin{equation}\label{eq3-8b}
\displaystyle{\left.\frac{\p\left(p^{n+1}-p^{n}\right)}{\p\textbf{n}}\right|_{\Gamma}=0,}
\end{equation}
\end{subequations}
which we solve with an iterative biconjugate gradient stabilized algorithm
(Saad 1996). In the horizontal directions, periodic boundary conditions are
imposed on the velocity and pressure, thus excluding a net horizontal
pressure gradient. For the simulations we will present here, 
this choice is consistent with the requirement of no mean horizontal flux
in a large but bounded container.  

We note that in the implicit solution of (\ref{eq3-5}), we apply the same biconjugate gradient stabilized solver used for the pressure to each component of $\tilde\textbf{u}=(\tilde u,\tilde v, \tilde w)$ separately. Thus only the diagonal terms of the diffusion operator are treated fully implicitly. The off-diagonal terms are treated quasi-implicitly in that the newest available values of $(\tilde u,\tilde v, \tilde w)$ are used in the evaluation of the cross derivatives. To ensure symmetry, we permute the order of solution for each component.

\section{Results: linear analysis}
\label{sec:linear}


\subsection{Floquet analysis}
\label{sec:floK}

In the absence of lateral boundaries, the equations are homogeneous 
in the horizontal coordinates and the solutions can be represented 
by a spatial Fourier transform:
\begin{equation}
\zeta(x,y,t) = \sum_\bk \hat{\zeta}(\bk,t) \eikx 
\label{eq:four}\end{equation}
The linear instability of the interface between two fluids is described by
(\ref{eq2-2})--(\ref{eq2-9}) linearized about a zero velocity field
and flat interface $\zeta=\langle\zeta\rangle$.
The linearized equations depend only on the wavenumber $k\equiv||\bk||$ 
of each wave and not on its orientation and hence the coefficient of 
$\eikx$ can be written as $\hat{\zeta}(k,t)$; additionally the dynamics of
each $\hat{\zeta}(k,t)$ is decoupled from the others.
Linear partial differential equations with constant coefficients have
solutions which are exponential or trigonometric in time.
For the Faraday instability $\hat{\zeta}(k,t)$ is instead governed by 
a system of linear partial differential equations with time-periodic 
coefficients, i.e. a Floquet problem, whose solutions are of the form
\begin{equation}\label{eq4-1}
\hat{\zeta}(k,t)=\mathrm{e}^{\left(\gamma+\mathrm{i}\alpha\omega \right)t}
\floq(k,t \textmd{ mod } T),
\end{equation}
where $T=2\upi/\omega$, $\gamma$ is real and $\alpha\in[0,1[$.
The Floquet modes,
\begin{equation}
\floq(k,t\textmd{ mod } T)=\sum_{n=-\infty}^{\infty} \floq_n(k)\mathrm{e}^{\mathrm{i}n\omega t},
\end{equation}
are not trigonometric,
but remain periodic with fundamental frequency $\omega$. 
Thus, the linearized behavior for a single mode is
\begin{equation}\label{eq4-4}
\zeta(x,y,t) = 
\eikx \mathrm{e}^{(\gamma +\mathrm{i}\alpha\omega) t}\sum_{n=-\infty}^{\infty} \floq_n(k)\mathrm{e}^{\mathrm{i}n\omega t}.
\label{eq:floK}\end{equation}
Analogous expressions hold for the velocity $\mathbf{u}$.


Equation (\ref{eq4-1}) shows that
if $\gamma$ is non-zero or $\alpha$ is irrational, the evolution of the interface
motion is not periodic. A non-zero $\gamma$ indicates that the motion grows or
decays according to the sign of $\gamma$. An irrational $\alpha$
yields a quasi-periodic evolution function.
For the Faraday instability, it can be shown that $\alpha$ can only take two values: 0 and
$1/2$. 
As the imposed acceleration $a$ is increased,
one encounters regions in the $(k,a)$ plane in which $\gamma>0$ for one dominant
temporal frequency, $j\omega/2$, where $j=1, 2, 3, ...$ (see
figure \ref{fig1}). Within each instability tongue, the amplitude of the
mode grows exponentially. These tongues are called harmonic if
$\alpha=0$ and subharmonic if $\alpha=1/2$. As $k$ is increased, one
encounters an alternating sequence of subharmonic and harmonic tongues, which
are bounded by neutral curves $(k,a_c(k))$ on which $\gamma=0$.  On the
neutral curves, the solutions are periodic:
\begin{subequations}\label{eq4-2}
\begin{eqnarray}
\hat{\zeta}(k,t)=\displaystyle{\sum_{n=-\infty}^{\infty} \floq_n(k)\mathrm{e}^{\mathrm{i}\left(n+\frac{1}{2}\right)\omega t},} && \qquad\textmd{subharmonic case,}
\label{eq4-2a}\\
\hat{\zeta}(k,t)=\displaystyle{\sum_{n=-\infty}^{\infty} \floq_n(k)\mathrm{e}^{\mathrm{i}n\omega t},} && \qquad\textmd{harmonic case.}
\label{eq4-2b}
\end{eqnarray}
\end{subequations}

\subsection{Computation of the neutral curves}
\label{sec:neutral}

We first compare our numerically calculated instability thresholds with those
found by \cite{KT1994} for the same parameter values.  The physical parameters
are $\rho_1 =519.933\kgpermc$ and $\mu_1 =3.908 \times 10^{-5}\Patimess$ for
the lower fluid and $\rho_2 =415.667\kgpermc$ and $\mu_{2} =3.124 \times
10^{-5}\Patimess$ for the upper fluid. The other parameters are $\sigma =2.181
\times 10^{-6}\Nperm$ and $g=9.8066\mperssq$.  The frequency of the forcing is
$\omega /2 \upi =100\Hz$ and thus its period is $T=0.01 $s.  The capillary length is defined 
as $l_c=\sqrt{\sigma/(|\rho_1-\rho_2|g)}$. The container
height is taken to be $5l_c=0.231\mm$, and the interface, when unperturbed, is equidistant from
the top and bottom boundaries. 
We consider several wavenumbers $k$ and set the $x$ dimension of the box in
each case to one expected wavelength $\lambda=2\upi/k$, i.e. to between
$0.074$ and $0.224\mm$, as listed in Table \ref{tab2}.  We can estimate the
importance of various physical effects for these parameters by defining
dimensionless quantities with length $k^{-1}$ and forcing period $T$.  The
Bond number $Bo= (kl_c)^{-2}=\left|\rho_1-\rho_2\right| g /(\sigma k^2)$ measures the
relative importance of gravitational to capillary effects and ranges between
$0.0649$ and $0.598$.  The Reynolds number $Re=\rho /(\mu k^2 T)$ is a
nondimensional measure of viscous damping and ranges between $0.184$ and $1.70$ for
both fluids.

We have computed the critical acceleration from our fluid-dynamical simulation
for the wavenumbers listed in Table \ref{tab2}. Initially, the interface is sinusoidal with
wavevector $\textbf{k}$ parallel to the $x$-axis and the velocity is zero.
Moreover, to ensure that the solution
corresponds initially to the linear solution, we require the amplitude of the
interface displacement to be small compared to $\lambda$.  In order to
maintain a roughly cubic mesh and a minimum $x$-resolution of about 50 grid
cells per wavelength, we vary the resolution in the $z$ direction
between 126 and 144 points. Since $\textbf{k}$ points along the $x$ direction,
$\zeta$ does not depend on $y$ (neither do the velocity nor the pressure) and
so the size of the domain and resolution in $y$ are arbitrary. The
acceleration $a$ is taken near $a_c(k)$, the expected critical acceleration
corresponding to each wavenumber. At the threshold, the flow undergoes a
pitchfork bifurcation. Since the growth rate is proportional to $a-a_c$ close
to the neutral curve, it is sufficient to find the growth rates for two values
of $a$ and to interpolate linearly between these points.

\begin{figure}
\begin{center}
\includegraphics[width=12cm,height=8cm]{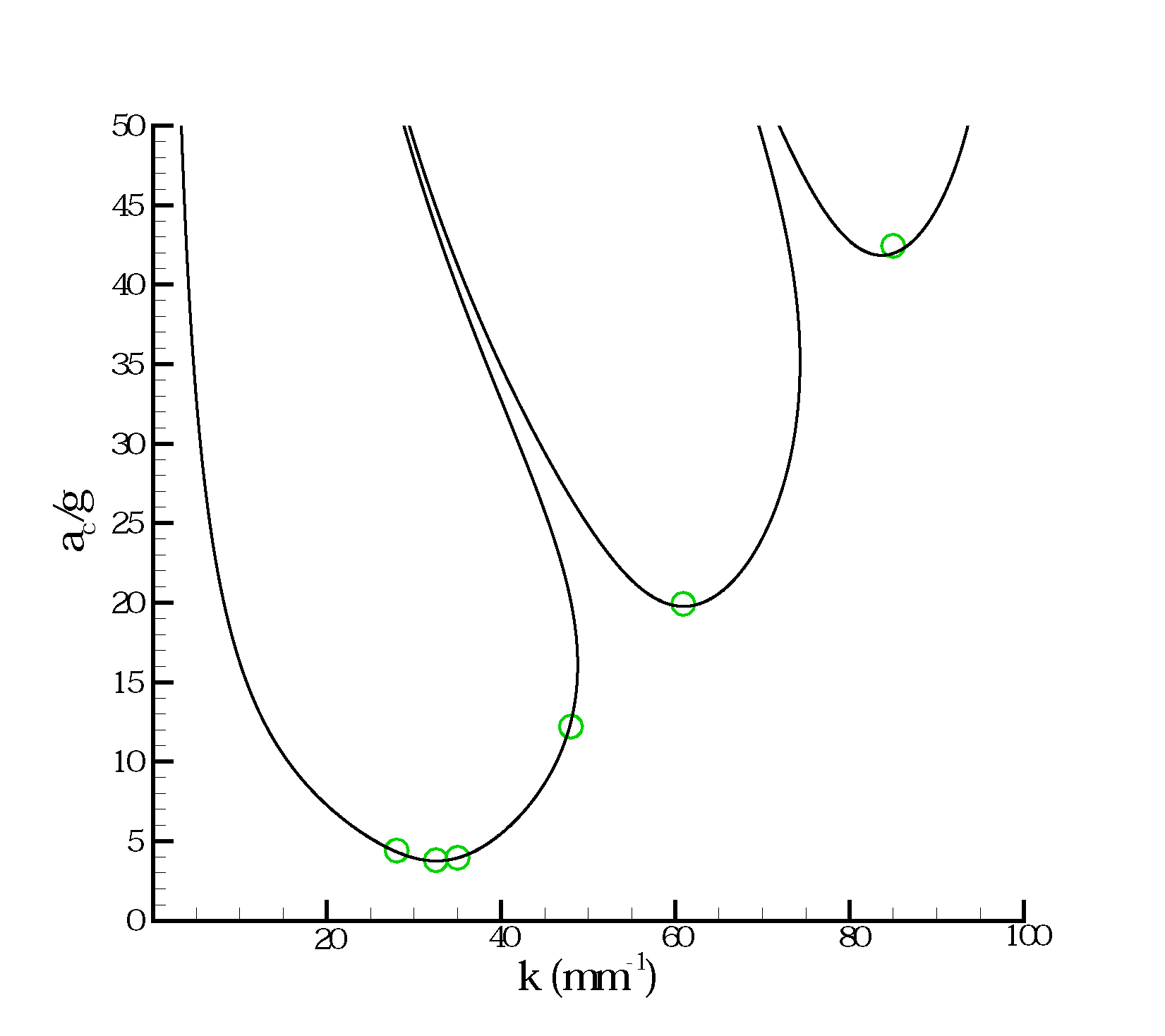}
\caption{Critical acceleration $a_c/g$ as a function of the wavenumber
  $k$. The solid curves represent the neutral curves obtained by
  \cite{KT1994}. The $a_c$ found with the simulation are 
indicated by the circles.}
\label{fig1}
\end{center}
\end{figure}

In figure \ref{fig1}, we plot the values of $a_c$ obtained from our
fluid-dynamical simulation for several values of $k$, along with the curves
$(k,a_c(k))$ obtained from the method of \cite{KT1994}.  Figure \ref{fig1}
shows that these thresholds are in good agreement, despite whatever
inaccuracies in $a_c$ are introduced by spatial discretization and linear
interpolation. The relative error in the critical acceleration at the
conditions previously stated is of the order of a few per cent as shown in
Table \ref{tab2}.

\begin{table}
\begin{center}
\begin{tabular}{@{}cccccc@{}}
$k$(\permm) & $\lambda$(\mm)& No. of gridpoints in $x/z$ & $ a_c/g$ (Theor.) &
  $a_c/g$ (Comp.) & Error(\%)\\
28 & 0.224 & 124/128 & 4.375  & 4.407  &0.7\\
32.5 & 0.193 & 96/128 & 3.777  & 3.800  &0.6\\
35 & 0.180 & 100/128 & 3.960  & 3.954  &-0.1\\
48 & 0.131 & 72/126 &  12.506 & 12.207 &-2.4\\
60.9 & 0.103 & 56/126 & 19.760 & 19.922 &0.8\\
85 & 0.074 & 48/144 & 41.953 & 42.358 &1.0\\
\end{tabular}
\end{center}
\caption{Comparison of the computed $a_c$ with Floquet theory for various 
wavenumbers $k$.}
\label{tab2}
\end{table} 
 
The results suggest that there is a $k$ below which calculation of the growth rates is not possible.
Some zones of the diagram are not accessible because a domain of width $2\upi/k$ necessarily accommodates all wavenumbers which are integer multiples of $k$ up to the resolution limit $\upi/k \Delta x$. The coefficients of the Fourier expansion of the initial condition differ slightly from zero due to finite-difference spatial approximations and if the growth rate of one of these
is greater than that of $k$ itself, then it will quickly come to dominate
$k$. This difficulty is exacerbated by the fact that several forcing periods
are required for $\gamma$ to stabilize. Then the amplitude, whose evolution
was expected to be almost periodic, starts to rise before the precise
determination of $\gamma$ is possible, for example in the range of $k$
between 0 and roughly $15\permm$. As we see in figure \ref{fig1},
the critical forcing is substantially lower for one of its multiples closer to
$32.5\permm$. The amplitude corresponding to this wavelength,
although initially negligible, increases and rapidly dominates the mode 
we wish to study, making the calculation of $\gamma$ unfeasible.
In contrast, for $k=48\permm$, the growth rate did not vary significantly after having reached a value near zero (relative fluctuations of about $0.1\%$ of the growth rate's limit value were recorded after the stabilization).

\subsection{Temporal profile of a mode}

We recall from section \ref{sec:floK} that the time dependence of a Floquet
mode is not sinusoidal. As a further validation, we can compare the 
results of our fluid-dynamical simulations to the entire temporal behavior 
over a period. This is a stronger validation than merely
predicting the threshold since it provides a comparison at every time instead
of once per period. 

\begin{figure}
\begin{center}
\includegraphics[height=6cm]{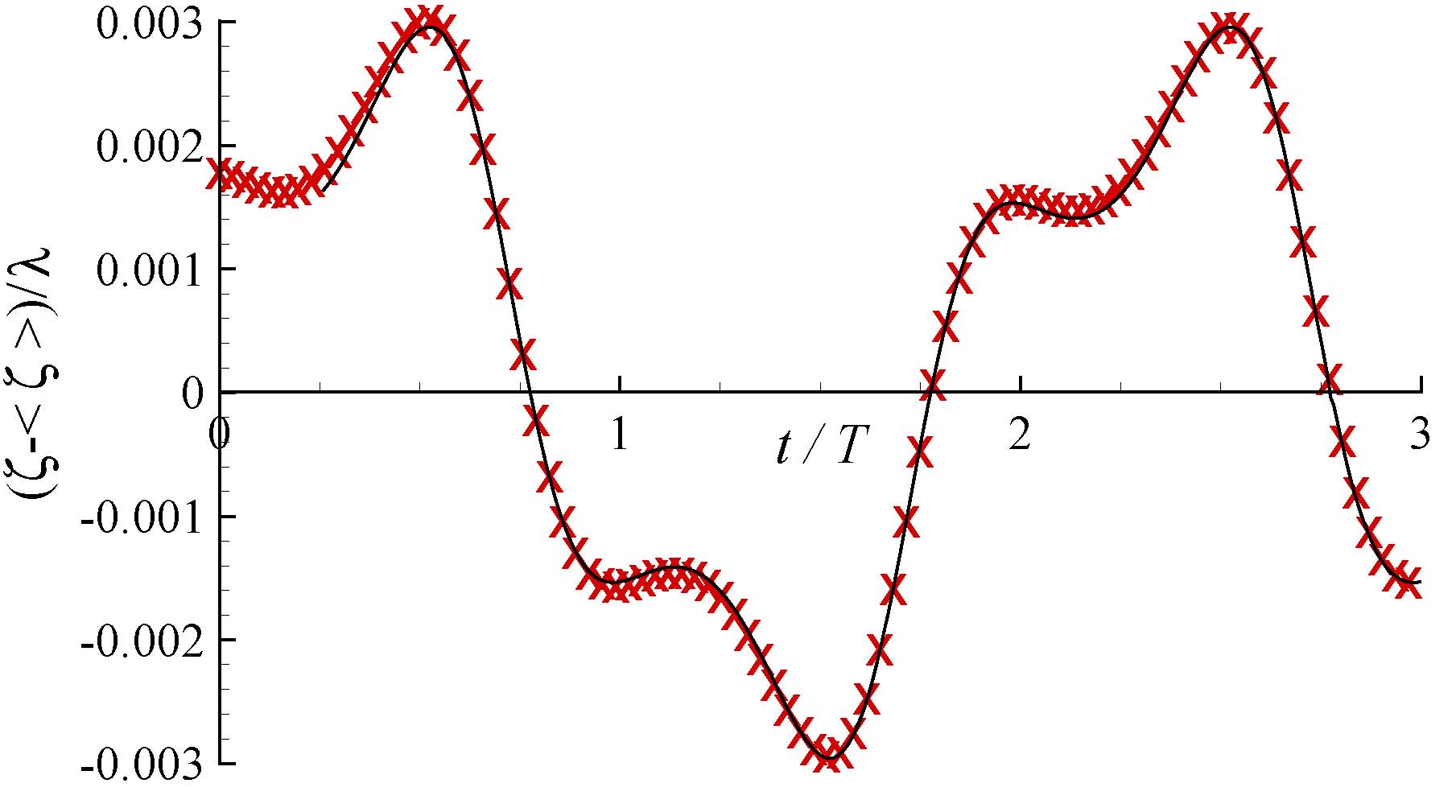}
\caption{Linear evolution of the surface height deviation $\zeta (t)-\langle\zeta\rangle$
for $k=48\permm$, in the 
first instability tongue. Our simulation results are plotted with symbols and
those derived from a Floquet analysis with the solid line. 
The height and time are nondimensionalized by the wavelength $\lambda=2\upi/k$ 
and forcing period $T$, respectively.}
\label{fig2}
\end{center}
\begin{center}
\includegraphics[height=6cm]{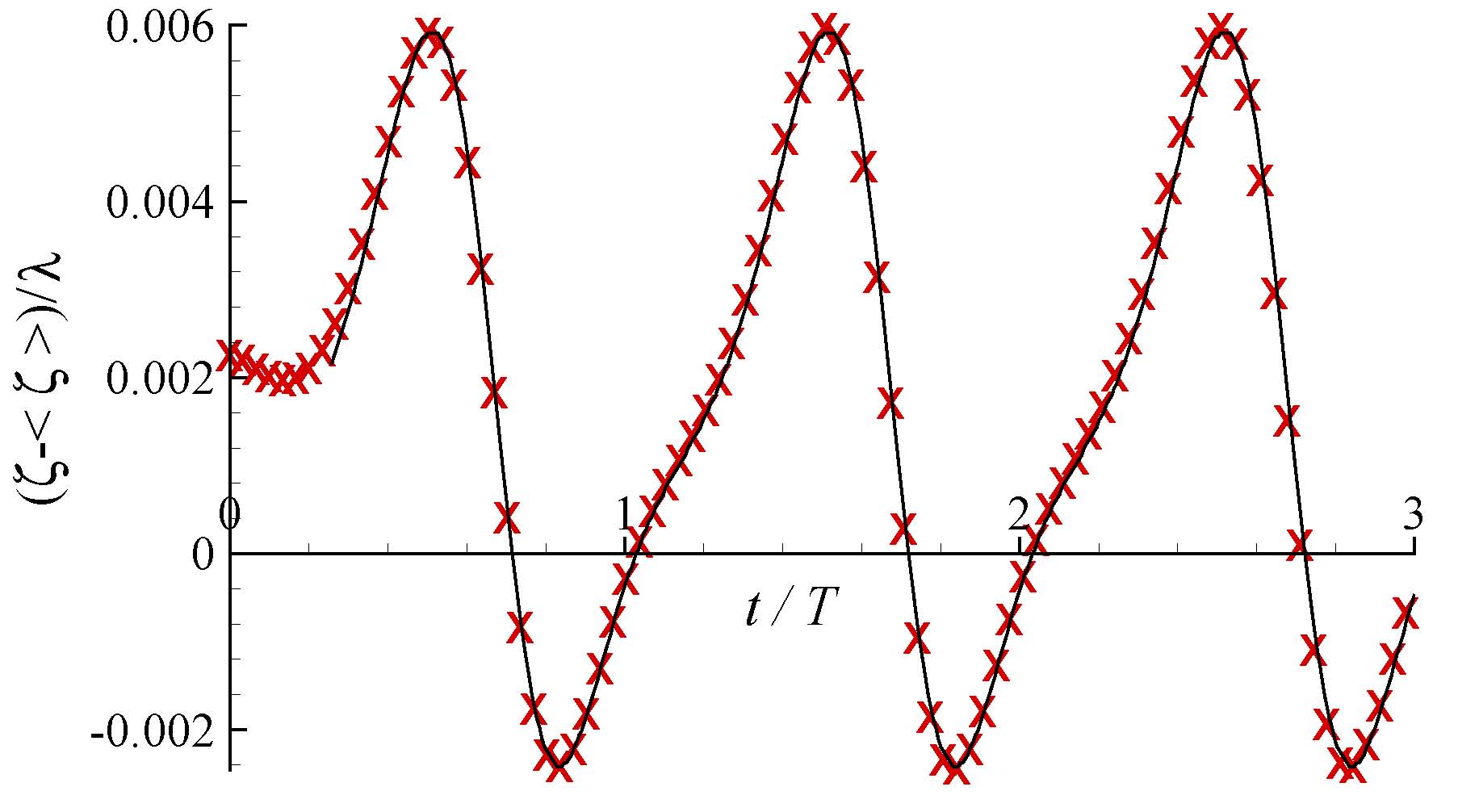}
\caption{Linear evolution of the surface height deviation $\zeta (t)-\langle\zeta\rangle$ for
  $k=60.9\permm$, 
in the second instability tongue. Same conventions used as figure \ref{fig2}.}
\label{fig3}
\end{center}
\begin{center}
\includegraphics[height=6cm]{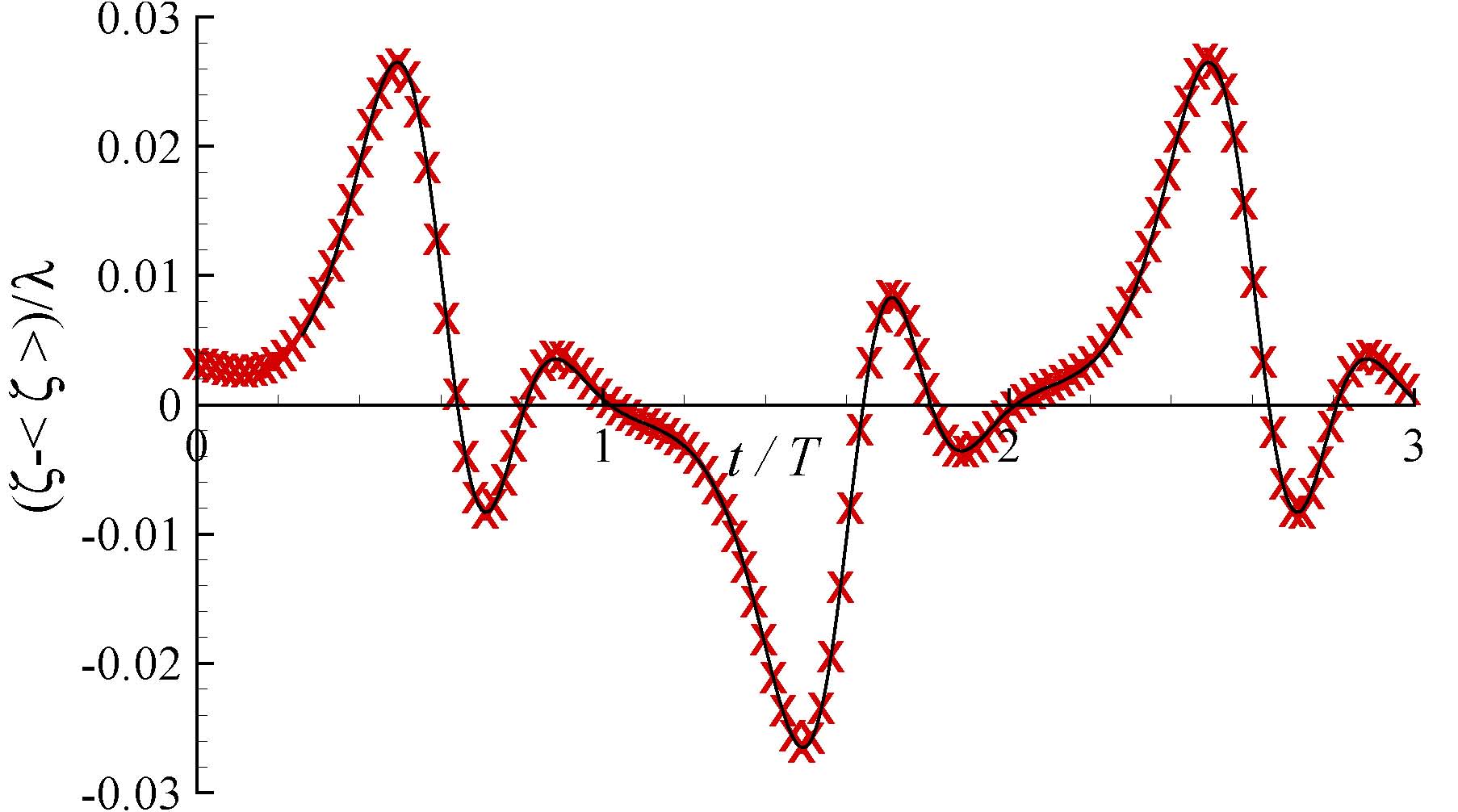}
\caption{Linear evolution of the surface height deviation 
$\zeta (t)-\langle\zeta\rangle$ for $k=85\permm$, 
in the third instability tongue. Same conventions used as figure \ref{fig2}.}
\label{fig4}
\end{center}
\end{figure}

In figures \ref{fig2}--\ref{fig4}, we plot the deviation 
$\zeta-\langle\zeta\rangle$ from the flat interface as a function of time 
at a fixed spatial location from our fluid-dynamical simulation, 
for values $k=48$, $60.9$ and $85\mm^{-1}$ belonging to the first three tongues.
On the same figures, we plot the behavior of (\ref{eq4-2}),
where the temporal coefficients $\floq_n(k)$ of the Floquet modes 
have been calculated by the method in \cite{KT1994}. 
The value of $a$ is set to the interpolated critical acceleration $a_c(k)$,
so the oscillations approximately retain their initial amplitude
as long as they remain small.
The comparisons in figures \ref{fig2}--\ref{fig4} show a nearly
perfect agreement. 
The differences observed initially, due to the phase difference between
the initial conditions and acceleration,
vanish remarkably quickly, in well under one period of oscillation of the
container. 

Figures \ref{fig2}--\ref{fig4} correspond to tongues
$j\omega/2$, with $j=$1, 2, 3, respectively, which show $j$ zero crossings
per forcing period $T$. Odd (even) values of $j$
correspond to subharmonic (harmonic) oscillations, with period $2T$ ($T$).
The temporal spectrum $\floq_n(k)$ becomes richer as $k$
increases, leading to increasingly more complex modes, as can be observed by
comparing figures \ref{fig2}--\ref{fig4}.  
This strong anharmonicity of the curves is due to the increasing contribution
of higher frequency trigonometric functions to the Floquet modes as $a$
increases.
The Floquet mode corresponding to $k_c =32.5\permm$, with the smallest value
of $a=a_c$, should be closer to trigonometric, with a fundamental frequency of
$\omega/2$.

\section{Results: nonlinear analysis}

In the full nonlinear evolution of the interface for $a>a_c$, the amplitude of the interface height
grows in time until nonlinear terms in (\ref{eq2-2})--(\ref{eq2-9}) 
become important. After that, the 
mode whose linear growth rate is maximal gives rise, via nonlinear resonances, 
to a
series of other discrete modes, selected according to the magnitudes and
orientations of their wavevector $\textbf{k}$. This selection is responsible
for the formation of patterns that will be the object of our further
validations. We seek to compare our calculations with the experimental results
of Kityk \etal~(2005, 2009) where quantitative data concerning the
Fourier spectrum $\hat{\zeta}(\textbf{k},t)$ are available for
squares and hexagons.  

We run our numerical simulations with the same experimental parameters as
\cite{KEMW2005}: $\omega/2\upi=12\Hz$ ($T=0.0833\s$), $\rho_1 =1346\kgpermc$,
$\mu_{1}=7.2\mPatimess$ for the lower fluid and $\rho_2 =949\kgpermc$,
$\mu_{2}=20\mPatimess$ for the upper fluid. The surface tension at the
interface is $\sigma=35\mNperm$, the total height of the vessel is $1.0\cm$ and
the mean height of the interface, the initial fill height of the heavy fluid,
is $\langle\zeta\rangle=1.6\mm$ (with some uncertainty; see below).  The
Floquet analysis for these parameters yields a critical wavelength of 
$\lambda_c =2\upi/k_c=13.2\mm$ and a critical acceleration of 
$a_c=25.8 \mperssq$.
Here, the Bond number defined in section \ref{sec:neutral} is $Bo=
\left|\rho_1-\rho_2\right| g /(\sigma k_c^2)=0.49$. 
The Reynolds number $Re=\rho /(\mu k_c^2 T)$ is
$Re_1=9.9$ and $Re_2=2.52$ for the lower and upper fluid, respectively.

Rather than starting from a sinusoidal interface, 
we chose to add two-dimensional white noise of small amplitude 
to $\langle\zeta\rangle$ to define the initial interface height 
$\zeta(x,y,t=0)$ in order to excite every mode allowed by the box's
horizontal dimensions and number of cells. It is thus possible to check that
the correct critical mode (that whose growth rate is maximal) emerges from the
linear dynamics. In order to reproduce the experimental results in a
computational domain of a minimal size, the dimensions in $x$ and $y$ of the
box must correspond to the periodicity and symmetries of the expected
pattern. The minimal required resolution along these directions has been found
to be between 40 and 50 cells per wavelength.  
The number of triangles used to represent the interface is
16 times the total number of horizontal gridpoints.
The number of cells in the $z$
direction is taken so that $\displaystyle{\min_{S'}\zeta(x,y,t)}$ is greater
than about the height of 3--5 cells.  The required vertical resolution thus
varies with the forcing amplitude. The initial velocity is taken to be zero.

\subsection{Square patterns}

\begin{figure}
\includegraphics[width=12cm]{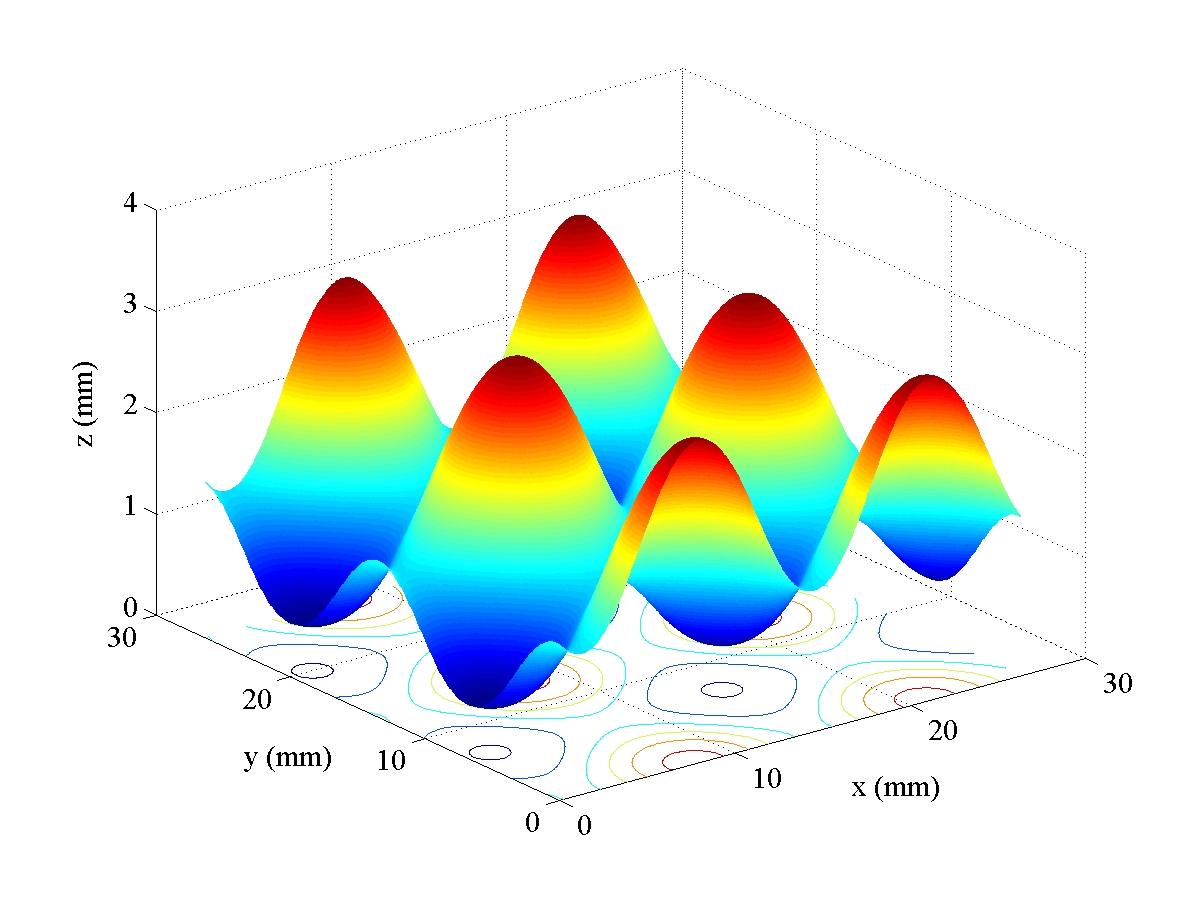}
\caption{Example of square pattern. Height of interface as a function of the
  horizontal coordinates, at the instant corresponding to first arrow of figure
  \ref{fig6}, when height is maximal.  Resolution in $x,y,z$ directions:
  $80\times80\times160$. Note that the vertical scale is stretched with 
respect to the horizontal scale. 
Each horizontal direction in the figure is twice 
that of the calculation domain.}
\label{fig5}
\includegraphics[width=12cm]{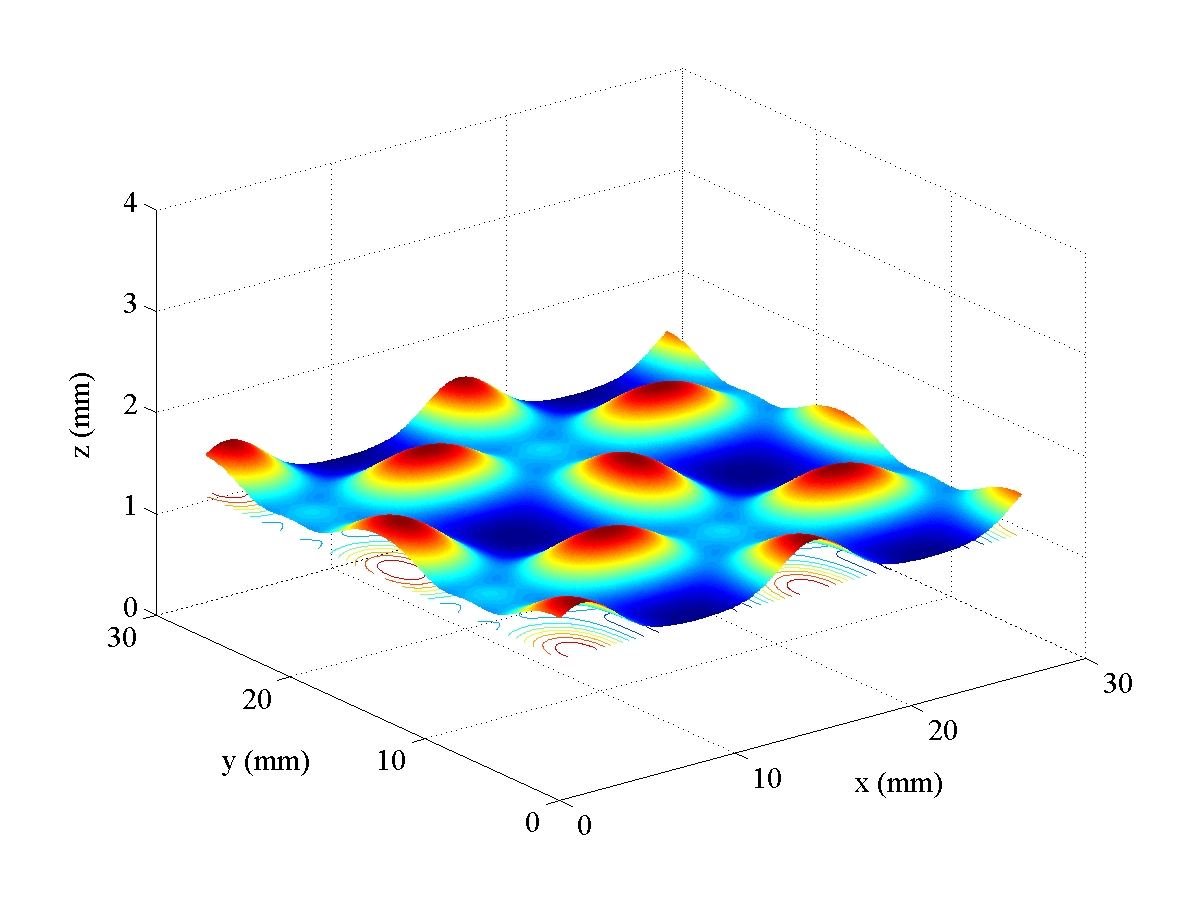}
\caption{Example of square pattern, 
at the instant corresponding to second arrow of figure \ref{fig6},
time $0.24\times 2T$ after figure \ref{fig5}.}
\label{fig5b}
\end{figure}

To compare with the experiment of \cite{KEMW2005} for their
square patterns, we choose the same forcing acceleration, $a=30.0\mperssq$. 
Our box has horizontal dimensions which we take both equal to
$2\upi/k_c$.  The timestep is $\Delta t = 2.78 \times 10^{-4}$\s.
Figures \ref{fig5} and \ref{fig5b} represent examples of the patterns obtained at saturation under these conditions and
are taken from the same simulation
at the two instants shown by the two arrows in figure \ref{fig6}. 
The symmetries characterizing the squares (reflections and
$\upi/2$ rotation invariance) are clear, showing a first qualitative agreement
with \cite{KEMW2005} where both structures were observed. 
The pattern oscillates subharmonically, at $2T$, where $T$ is 
the forcing period.
Figure \ref{fig5} 
is taken when the interface attains its maximum height, while figure 
\ref{fig5b} is taken at a time $0.24\times2T$ later. 
At this later time, we observe the dominance of a higher wavenumber, 
which will be discussed below.

\begin{figure}
\begin{center}
\includegraphics[height=8cm]{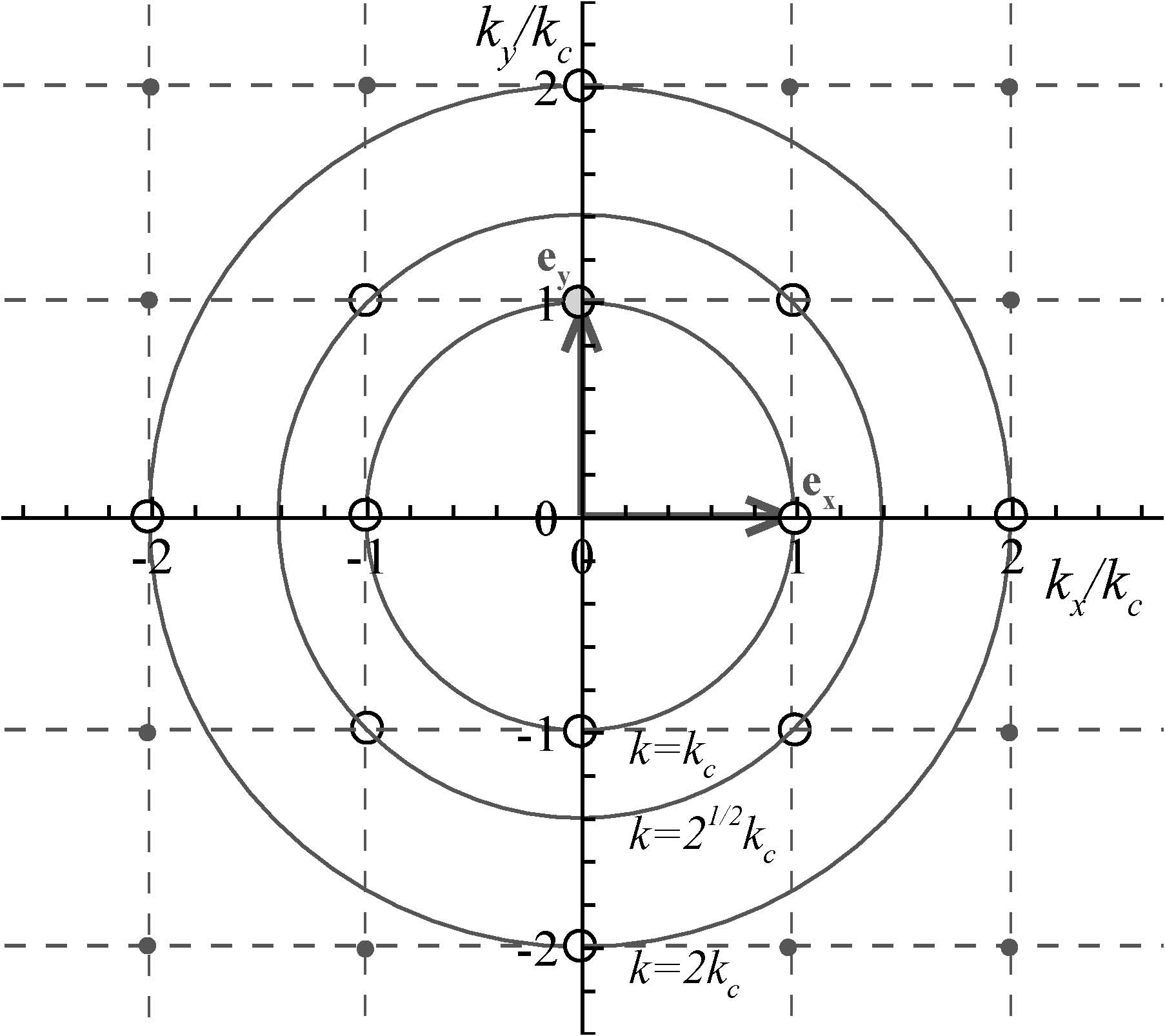}
\caption{Lattice formed by the spatial modes comprising a square
pattern. The principal modes, with wavenumbers
$k_c$, $2k_c$ and $\sqrt{2}\,k_c$,
whose evolution will be studied in figures \ref{fig6} and \ref{fig7} are
indicated by hollow black circles.}
\label{spsq}
\end{center}
\end{figure}

\begin{figure}
\begin{center}
\includegraphics[width=12cm]{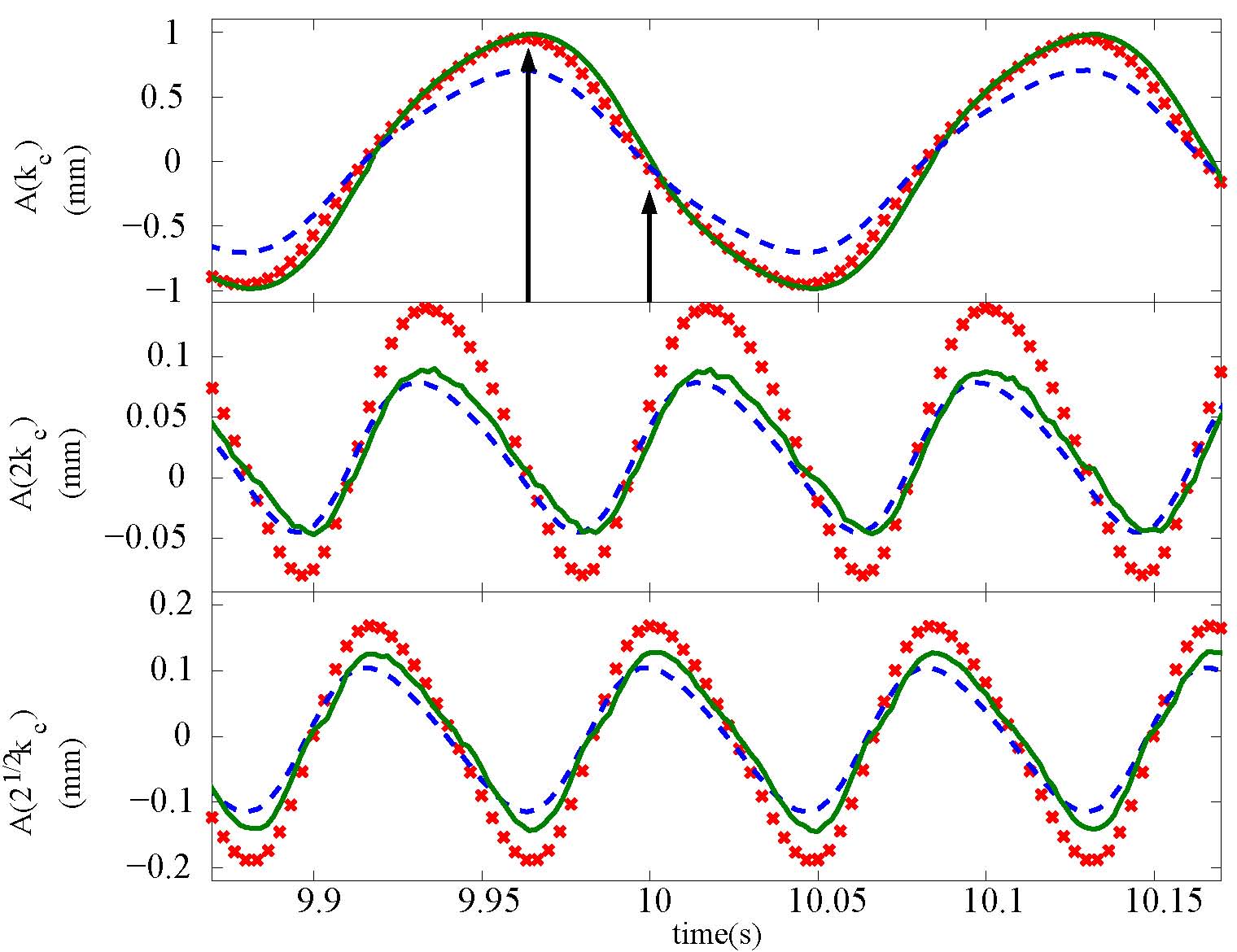}
\caption{Temporal evolution of the amplitudes of the spatial modes with
  wavenumbers $k_c$, $2k_c$ and $\sqrt{2}k_c$. Solid curves represent 
the experimental results of \cite{KEMW2009}, dashed
  curves and crosses represent numerical results for
  $\langle\zeta\rangle=1.6\mm$ and $\langle\zeta\rangle=1.7\mm$, respectively.
  Resolution in $x,y,z$ directions: $80\times 80\times 160$.  Arrows, from
  left to right, show the time at which figures \ref{fig5} and \ref{fig5b}
  have been plotted.}
\label{fig6}
\end{center}
\begin{center}
\includegraphics[width=12cm]{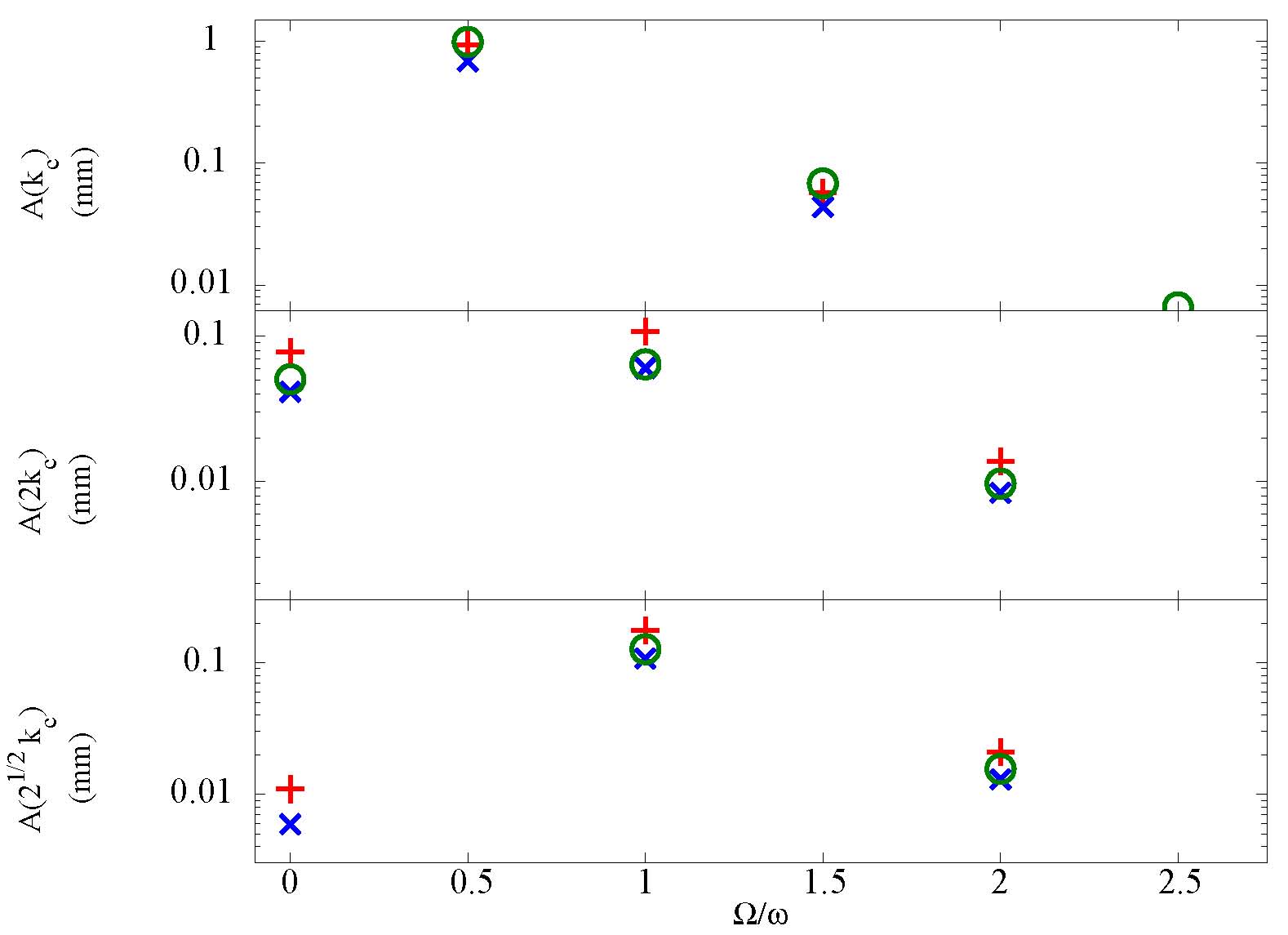}
\vspace*{-0.2cm}
\caption{Temporal Fourier transform of the amplitudes in figure \ref{fig6}.
Circles indicate experimental results of \cite{KEMW2009}, while crosses 
and plus signs indicate numerical data with
  $\langle\zeta\rangle=1.6\mm$ and $\langle\zeta\rangle=1.7\mm$ respectively.}
\label{fig7}
\end{center}
\end{figure}

Further quantitative investigations of the patterns involve the spatial
Fourier transform of the interface height. In the case of square patterns, the
distribution of the spatial modes is shown in figure \ref{spsq}. 
The modes with non-negligible amplitude are 
$\pm k_c \textbf{e}_x$ and $\pm k_c \textbf{e}_y$, with $|\textbf{k}|=k_c$ 
and amplitude $A(k_c)$;
$\pm 2k_c \textbf{e}_x$ and $\pm 2k_c \textbf{e}_y$, with $|\textbf{k}|=2k_c$ 
and amplitude $A(2k_c)$;
and $k_c(\pm\textbf{e}_x\pm\textbf{e}_y)$, with $|\textbf{k}|=\sqrt{2}k_c$ 
and amplitude $A(\sqrt{2}k_c)$.
(For a square pattern, the amplitude of each mode is identical to 
that of each of its images through rotation by any integer multiple of $\pi/2$.)
The interface height is written as:
\begin{eqnarray}
\zeta(\mathbf{x},t)=\langle \zeta \rangle 
&+& A(k_c,t)\:\sum_{j=1}^4 \mathrm{e}^{\mathrm{i} k_c \be_j \cdot \mathbf{x}}
+ A(2k_c,t)\:\sum_{j=1}^4 \mathrm{e}^{\mathrm{i} 2 k_c \be_j \cdot \mathbf{x}}  \nonumber \\
&+& A(\sqrt{2}k_c,t)\:\sum_{j=1}^4 \mathrm{e}^{\mathrm{i} \sqrt{2}k_c\be_j^\prime\cdot\mathbf{x}}+ \mbox{higher order terms,}
\end{eqnarray}
where $\be_j\equiv\be_x \cos(\pi j/2) + \be_y \sin(\pi j/2)$
and $\be_j^\prime\equiv\be_x \cos(\pi/4 + \pi j/2) + \be_y \sin(\pi/4 + \pi j/2)$
for $j=1,\ldots 4$.
We have chosen this notation, rather than $\hat{\zeta}(k,t)$ as used in
equation (\ref{eq:four}), to facilitate 
comparison with Kityk \etal~(2005, 2009).

We have compared the evolution of the three principal spatial modes (figure \ref{fig6}) and their
temporal Fourier transform (figure \ref{fig7}) with the experimental results
(Kityk \etal~2005).  Here we turn the reader's attention to the recent 
erratum by \cite{KEMW2009} for correct quantitative comparisons of the spectra. 
\begin{figure}
\begin{center}
\includegraphics[width=12cm]{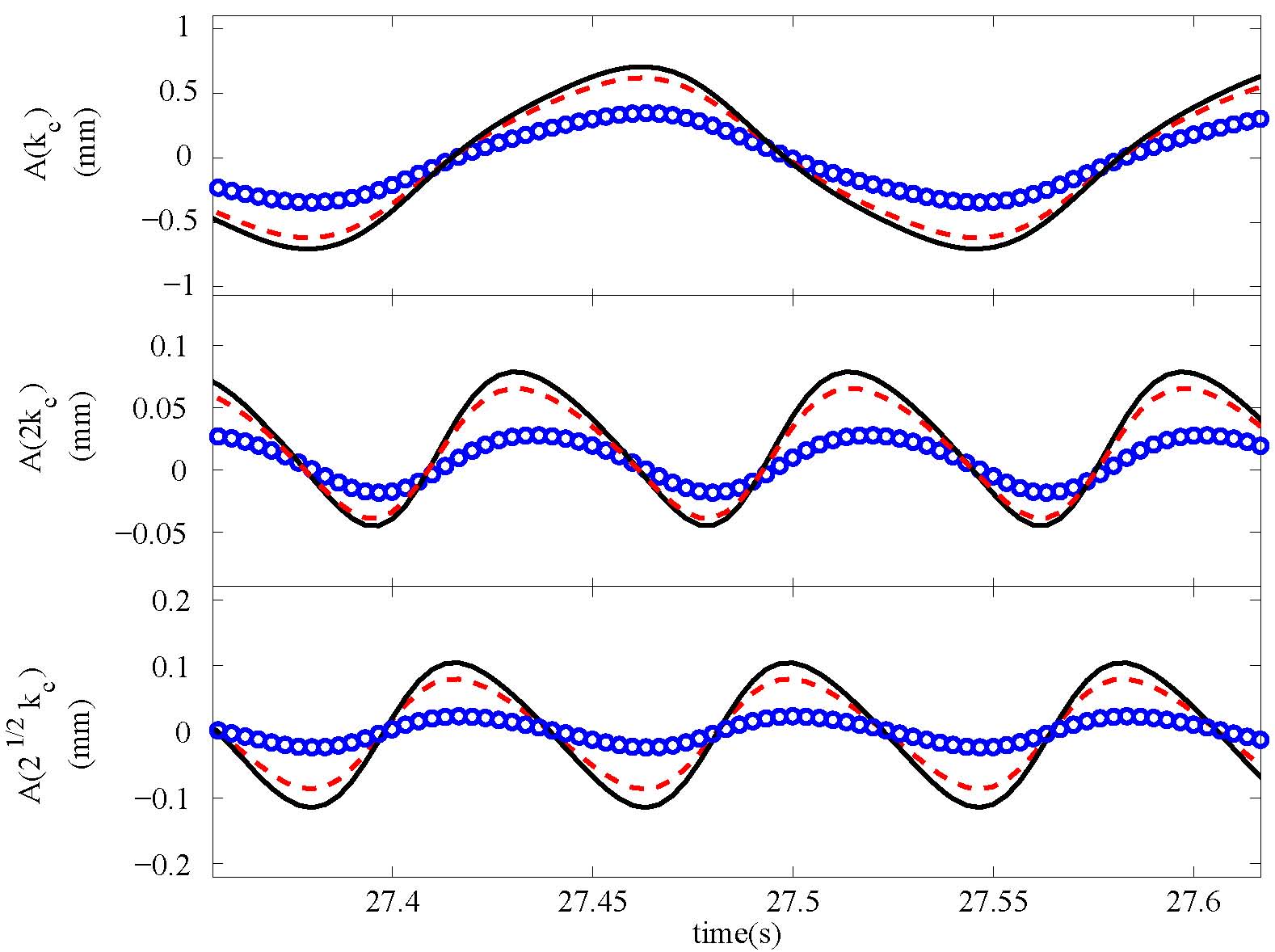}
\vspace*{-0.2cm}
\caption{Temporal evolution of the amplitudes of the spatial modes with wavenumbers 
$k_c$, $2k_c$ and $\sqrt{2}k_c$ for square
  patterns. Study of the convergence with three different spatial
  resolutions. Circles indicate a resolution (in $x,y,z$ directions) of
  $20\times20\times40$, dashed curves $40\times40\times80$ and continuous
  curves $80\times80\times160$. 
The timestep is the same, $\Delta t = 2.78 \times 10^{-4}\s$, for all curves shown.}
\label{fig-conv}
\end{center}
\end{figure}

Figure \ref{fig6} compares the experimental evolution of each spatial
wavenumber to numerical calcuations for two different mean heights
$\langle\zeta\rangle=1.6\mm$ and $\langle\zeta\rangle=1.7\mm$.
Our calculations
show that the results depend strongly on $\langle\zeta\rangle$,
which is the initial fill height of the heavy fluid.
Our discussions with
Kityk and Wagner (A. Kityk \& C. Wagner, private communication) indicate that this is true as well
in the experiments, and also that 0.1 mm is within the experimental uncertainty
for their mean height.  Thus we chose to vary $\langle\zeta\rangle$, in preference to
other parameters, in order to check
whether the range of amplitudes caused by experimental uncertainties includes
those obtained numerically.

The main features found in \cite{KEMW2009} are recognized
in figure \ref{fig6}. In particular, both the fundamental periodicity of each
mode (harmonic or subharmonic) and the form of each numerical curve in figure
\ref{fig6} are very similar to the experimental data. The amplitudes and
the phases are also quite close. Most of the experimental amplitudes are
bracketed by the numerical ones. 
Thus, they lie in the interval of amplitudes
allowed by the range of uncertainties which is surely underestimated since
only the uncertainty in $\langle\zeta\rangle$ has been taken into account.
$A(k_c)$ crosses zero at times different from
the two higher wavenumbers, $A(2k_c)$ and $A(\sqrt{2}k_c)$.
At these instants, the higher wavenumbers dominate the pattern.
In particular, the pattern of figure \ref{fig5b}, taken near the second 
arrow in figure \ref{fig6}, when $A(k_c)$ is low, contains more 
peaks than that of figure \ref{fig5}, taken when $A(k_c)$ is high.
The large ratio between the amplitude of
$k_c$ and the others makes this phenomenon very short-lived.  

Figure \ref{fig7} shows the temporal Fourier decomposition of the curves in
figure \ref{fig6}.  These spectra for the experiment \cite{KEMW2009} and for the computation
are quite similar too.  All of the square patterns that we have observed, once saturation is attained, remain so for the entire duration of the calculation.

We present a brief numerical grid convergence study in figure \ref{fig-conv}.
All qualitative features, such as the square symmetry, were observed with each
of the three resolutions chosen, despite the coarseness of the
$20\times20\times40$ and $40\times40\times80$ grids.  With increasing
resolution, the principal spatial modes converge to the experimental curves
shown in figure \ref{fig6}, with only a small difference between the curves
with the two highest resolutions.  The order of numerical convergence of the
maximum and minimum of the amplitudes of each of the three modes in figure
\ref{fig-conv} shows that the convergence is between first and second order,
which is expected to be the case with the immersed-boundary method.  In
particular, we would expect that a further doubling of the resolution would
change the results by at most 4 \% for the principal $k_c$ mode.

\subsection{Hexagonal patterns}

\begin{figure}
\begin{center}
\includegraphics[width=12cm]{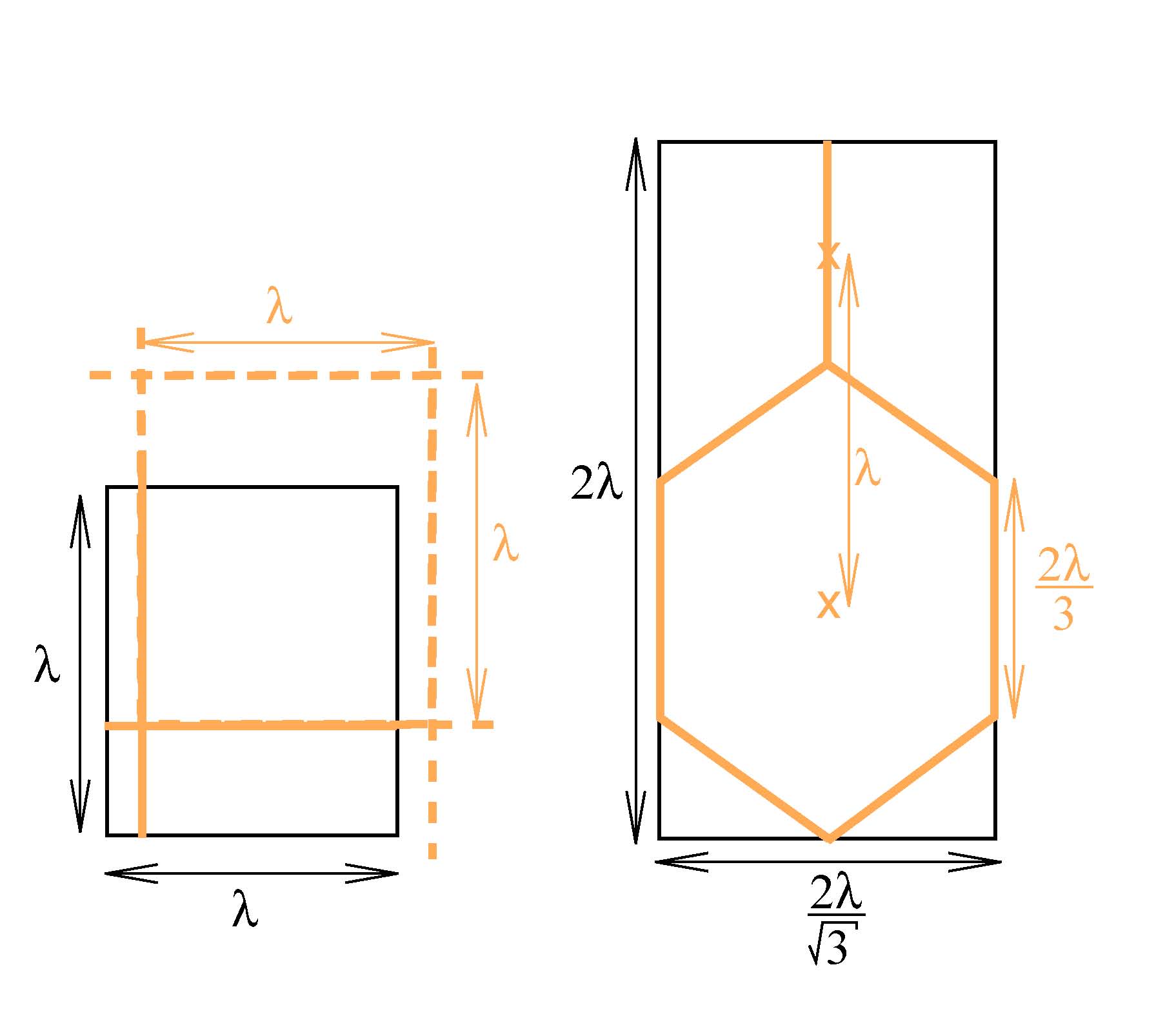}
\caption{Boxes supporting the periodic patterns in the square and hexagonal cases. In black, the borders of the box. Light lines, pattern contained by each box; $\lambda=2\upi/k_c$.}
\label{fig8}
\end{center}
\end{figure}

\begin{figure}
\begin{center}
\includegraphics[height=8cm]{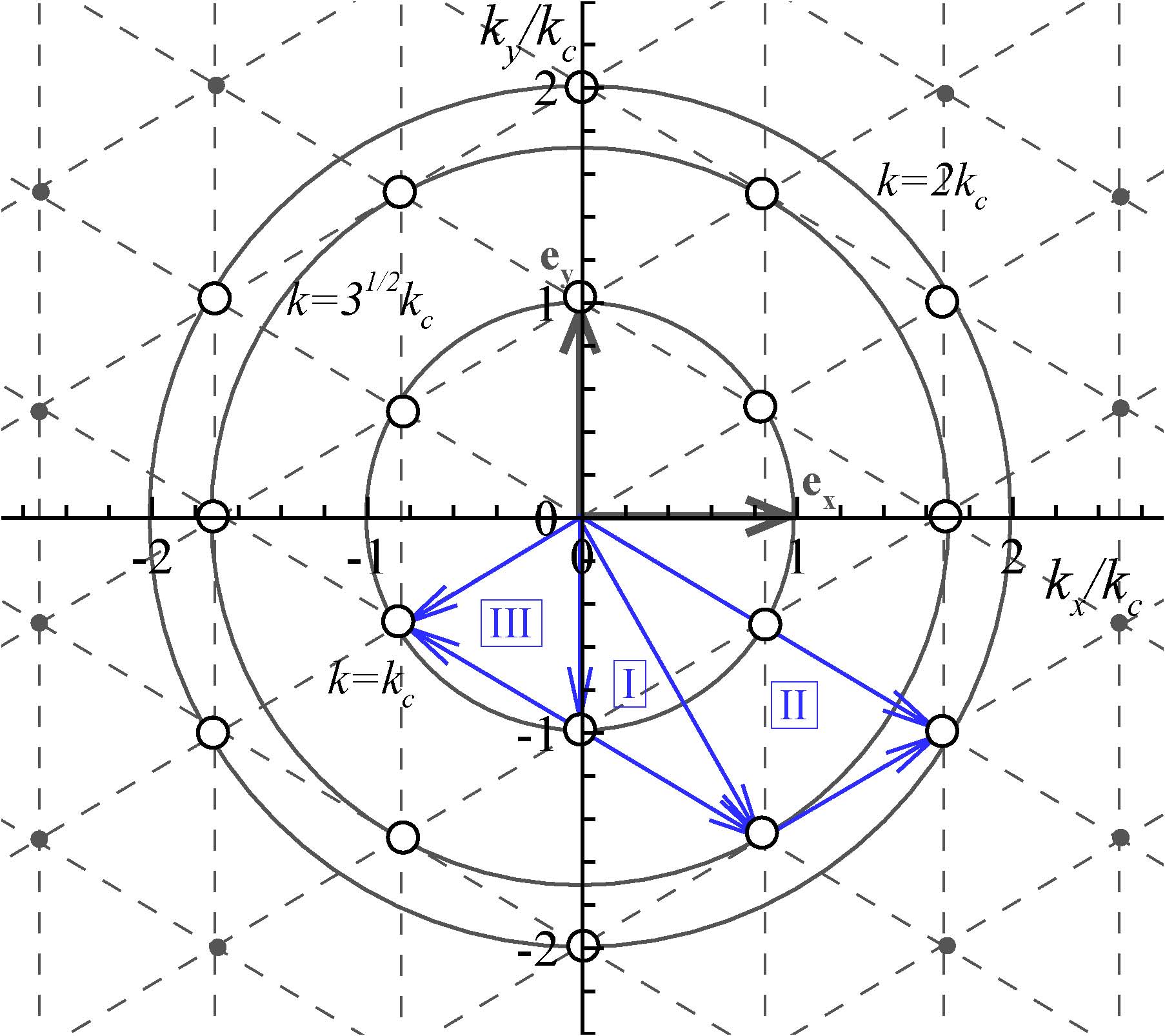}
\caption{Lattice formed by the spatial modes comprising a hexagonal
  pattern. The principal modes, with wavenumbers $k_c$, $2k_c$ and 
$\sqrt{3}\,k_c$, involved in later quantitative investigations are 
indicated by hollow black circles. The labelled triangles illustrate resonance mechanisms 
leading to harmonic contributions to higher wavenumbers.} 
\label{sphex}
\end{center}
\end{figure}

When the amplitude of the forcing acceleration $a$ is further increased, the
modes can reorganize. The symmetries change and, in the experiments of
\cite{KEMW2005}, the initial square pattern becomes hexagonal. Though $k_c$
remains constant, the horizontal dimensions of the minimal computational box necessary to
support the periodic pattern must change too. These dimensions become
$4\upi/k_c$ in $y$ and $4\upi/(\sqrt{3}k_c)$ in $x$, as shown in figure
\ref{fig8}.  The wavevector lattice for hexagonal patterns is shown in figure
\ref{sphex}.  The principal modes are again of three amplitudes: $k_c$,
$2k_c$ and $\sqrt{3}k_c$.  When a pattern is hexagonal, a mode will have the
same amplitude and temporal behaviour as each of its images through rotations
by any integer multiple of $\upi/3$. The interface height is thus
\begin{eqnarray}
\zeta(\mathbf{x},t)=\langle \zeta \rangle 
&+& A(k_c,t)\sum_{j=1}^6 \mathrm{e}^{\mathrm{i} k_c \be_j \cdot \mathbf{x}}
+ A( 2 k_c,t)\sum_{j=1}^{6} \mathrm{e}^{\mathrm{i} 2 k_c \be_j \cdot \mathbf{x}} \nonumber\\
&+& A(\sqrt{3}k_c,t)\sum_{j=1}^6 \mathrm{e}^{\mathrm{i} \sqrt{3}k_c \be_j^\prime \cdot \mathbf{x}} + \mbox{higher order terms,}
\end{eqnarray}
where $\be_j \equiv\be_x \cos(\pi j/3) + \be_y \sin(\pi j/3)$ 
and $\be_j^\prime \equiv \be_x \cos(\pi/6 + \pi j/3) + \be_y \sin(\pi/6 + \pi j/3)$.
for $j=1,\ldots, 6$.

Our simulations are carried out at acceleration $a=38.0\mperssq$ and 
mean height $\langle\zeta\rangle=1.6\mm$.
We have used two different initial conditions: a rectangular pattern,
and also white noise, as in our previous simulations of the square patterns.
In both cases, hexagons emerge and saturate. The results shown below are those that emerge from the white noise.
The time step varies during the calculation, depending on the viscous diffusion limit and the CFL. The spatial resolution is 58x100x180 in the $x$, $y$, $z$, directions, respectively.

\begin{figure}
\begin{center}
\vspace*{-1cm}
\includegraphics[width=12cm]{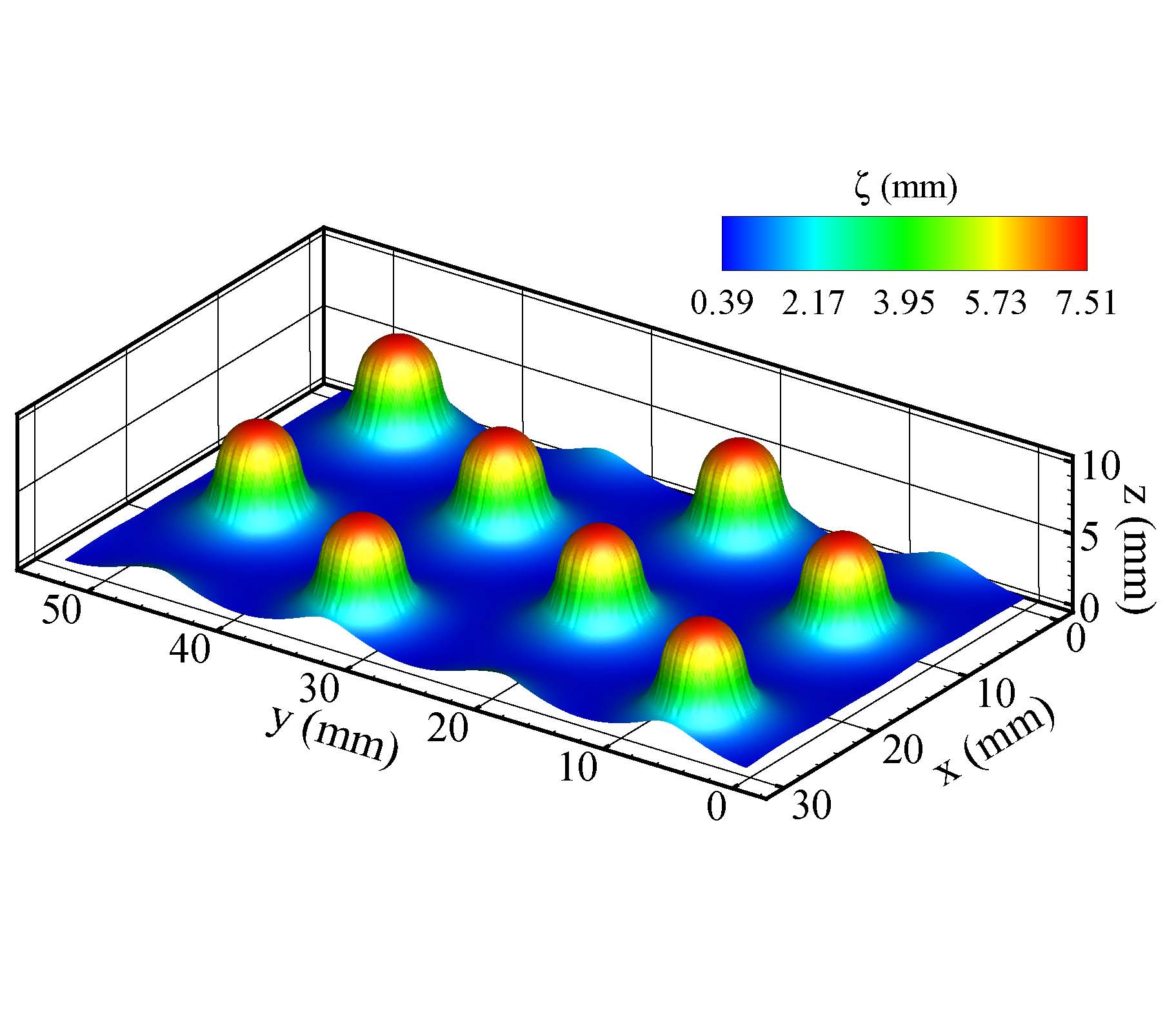}
\vspace*{-2cm}
\caption{Snapshot of hexagonal pattern, taken when height of the interface peaks is maximal. Each horizontal direction is twice that of the calculation domain.}
\label{fig9}
\end{center}
\begin{center}
\includegraphics[width=12cm]{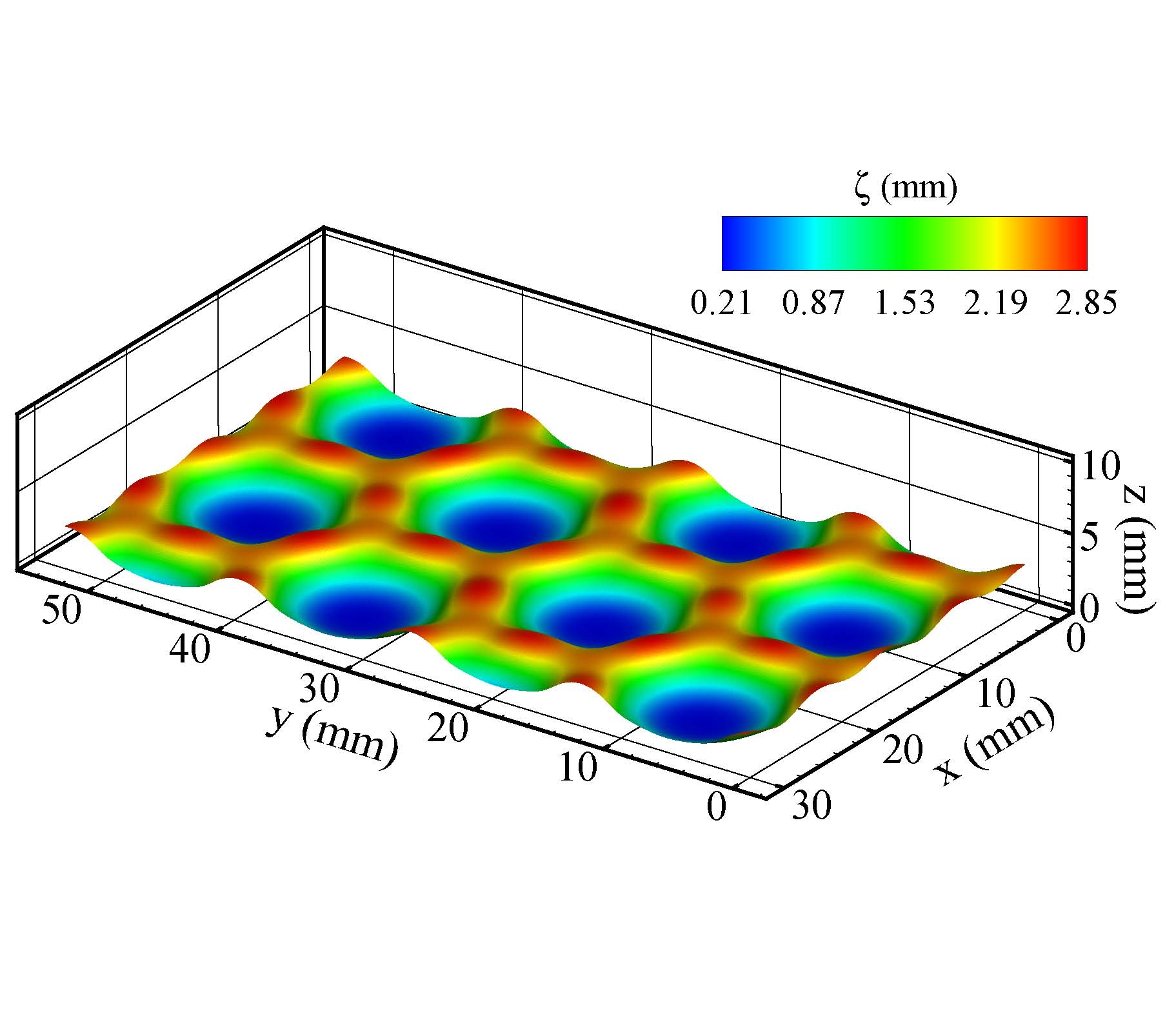}
\vspace*{-2cm}
\caption{Snapshot of hexagonal pattern taken $t=0.3\times 2T$ 
after instant of maximal interface height.}
\label{fig10}
\end{center}
\end{figure}

\begin{figure}
\begin{center}
\vspace*{-1cm}
\includegraphics[width=12cm]{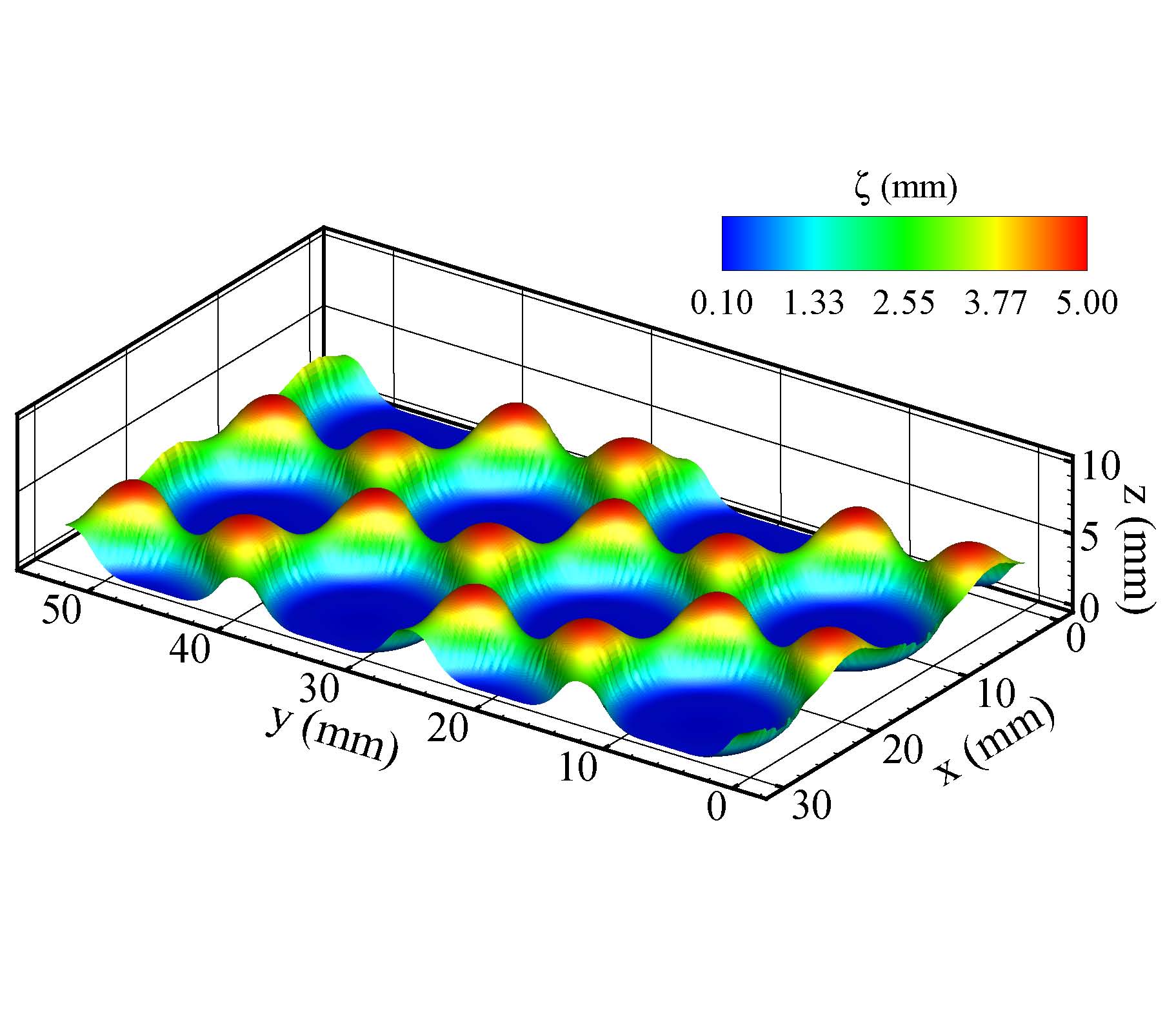}
\vspace*{-2cm}
\caption{Snapshot of hexagonal pattern taken $t=0.48 \times 2T$ 
after instant of maximal interface height.}
\label{fig10a}
\end{center}
\begin{center}
\includegraphics[width=12cm]{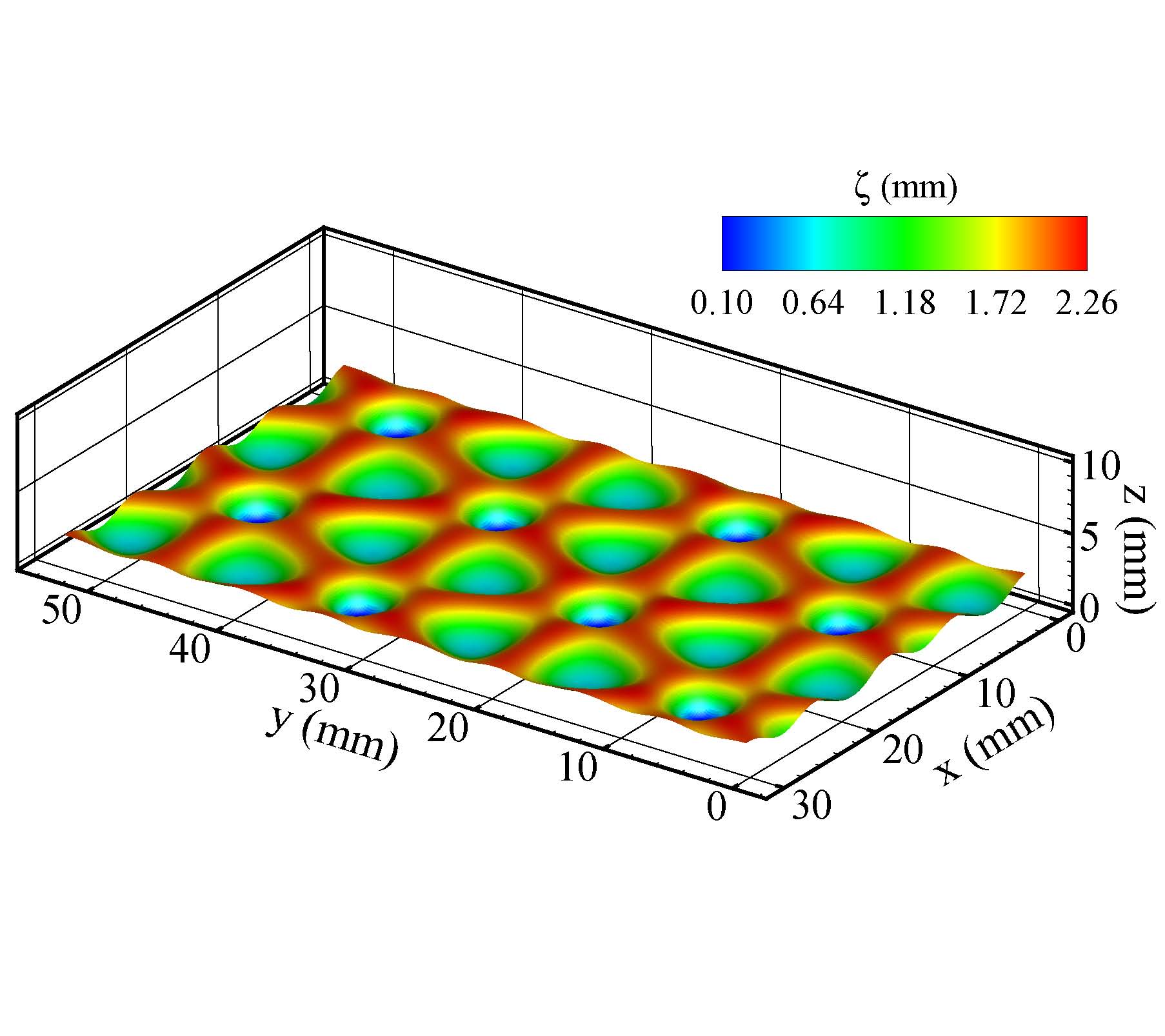}
\vspace*{-2cm}
\caption{Snapshot of hexagonal pattern taken $t=0.68 \times 2T$ 
after instant of maximal interface height.}
\label{fig10b}
\end{center}
\end{figure}

\begin{figure}
\begin{center}
\includegraphics[width=12cm]{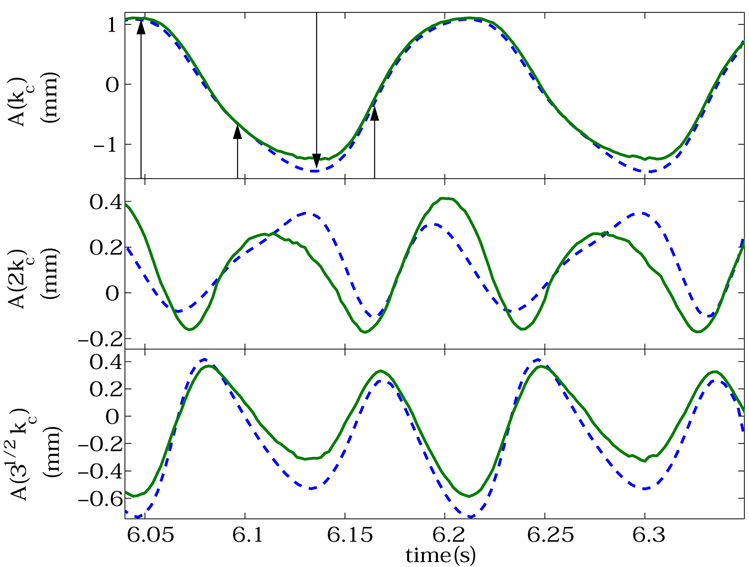}
\caption{Temporal evolution of the amplitudes of the spatial modes with
  wavenumbers $k_c$, $2k_c$ and $\sqrt{3}k_c$.
Solid curves represent experimental results (A. Kityk \& C. Wagner, private communication) at $a\approx 38.5\mperssq$. Dashed curves represent the simulation 
for $a=38.0\mperssq$ at resolution 
(in $x,y,z$ directions) of $58\times100\times180$. Arrows indicate times corresponding to figures \ref{fig9}--\ref{fig10b}.}
\label{fig11}
\end{center}
\begin{center}
\includegraphics[width=12cm]{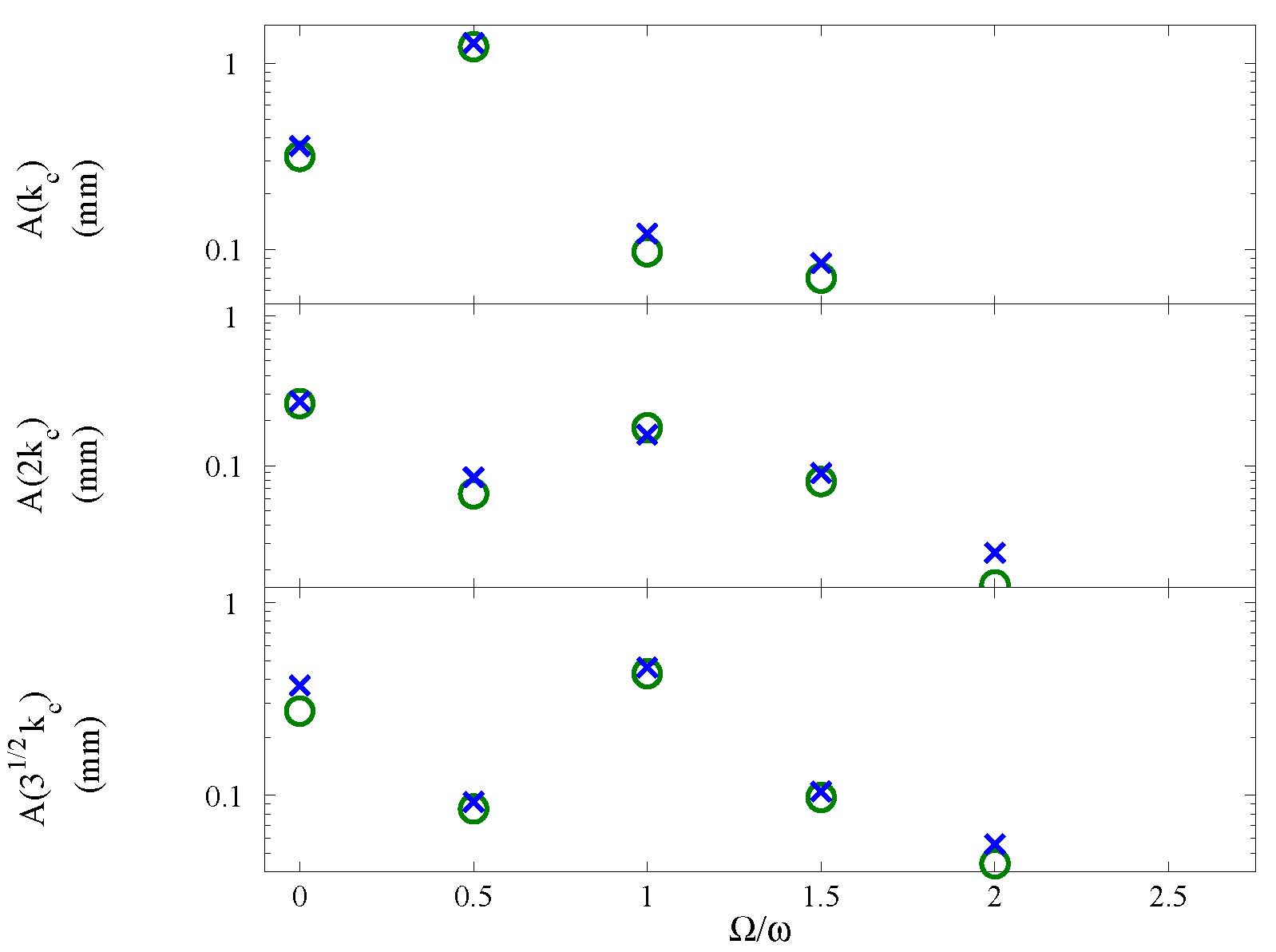}
\caption{Temporal Fourier transform of the amplitudes in figure \ref{fig11}. 
Circles represent experimental results (A. Kityk \& C. Wagner, private communication)
for $a\approx 38.5\mperssq$.
Crosses represent numerical results for $a=38.0\mperssq$
at resolution $58\times100\times180$.}
\label{fig12}
\end{center}
\end{figure}

In figures \ref{fig9}--\ref{fig10b}, we show visualizations of the patterns at four instances in time.  A movie of the temporal evolution 
of the hexagon pattern over one subharmonic oscillation is available in the online version of this article.
The $\upi/3$ rotational symmetry confirms that the rectangular numerical grid
does not forbid the formation of hexagonal patterns, which are not aligned
with this grid.  
The patterns reproduce several prominent features from 
the visual observations of hexagons in the experiments.
For example, one can observe in figures \ref{fig9} and \ref{fig10a} 
the up and down hexagons shown in the experimental snapshots (figure 10
of Kityk \textit{et al.} 2005).  
The pattern in figure \ref{fig10b}, when the surface elevation is minimal, 
is dominated by wavenumbers higher than $k_c$, as is also the case 
in figure 10 of \cite{KEMW2005}.
This is reflected by the disappearance of $A(k_c)$ and the
resulting dominance of $A(2k_c)$ and $A(\sqrt{3}k_c)$ at the 
corresponding instant in the spectral timeseries of figure \ref{fig11}.
This apparent wavenumber increase is analogous to that which occurs
for the squares, shown in figures \ref{fig5b} and \ref{fig6}.


The spectra from experiments and simulations are
represented in figures \ref{fig11} and \ref{fig12}.  Given that experimental
uncertainties concerning the hexagons are greater than for the squares (A. Kityk
\& C. Wagner, private communication), the agreement is remarkable. The principal
mode is well reproduced while the other two modes show rough agreement.  It is
striking that, in contrast to square patterns, every wavevector is a
superposition of harmonic and subharmonic temporal modes, so that 
each has temporal period $2T$.  This phenomenon was
explained by \cite{KEMW2005} as a spatio-temporal resonance as follows.  In
the case of the square lattice, two critical subharmonic modes (e.g.~$k_c
\textbf{e}_x$ and $k_c \textbf{e}_y$) interact to yield a higher wavenumber
harmonic mode (e.g.~$k_c (\textbf{e}_x+\textbf{e}_y)$).  In the hexagonal
case, two critical subharmonic modes (e.g.~$-k_c \textbf{e}_y$ and $k_c
(\sqrt{3}\textbf{e}_x-\textbf{e}_y)/2$) interact to yield a higher wavenumber
harmonic mode ($k_c(\sqrt{3}\textbf{e}_x-3\textbf{e}_y)/2$), as in triangle I
of figure \ref{sphex}.  Further interaction of this mode with a critical
subharmonic mode ($k_c(\sqrt{3}\textbf{e}_x+\textbf{e}_y)/2$) yields
subharmonic contributions to the higher spatial wavenumber mode
($k_c(\sqrt{3}\textbf{e}_x-\textbf{e}_y)$), as shown in triangle II. Other
quadratic interactions between critical subharmonic modes can contribute to a
third harmonic mode of wavenumber $k_c$ (triangle III).

\begin{figure}
\includegraphics[width=12cm]{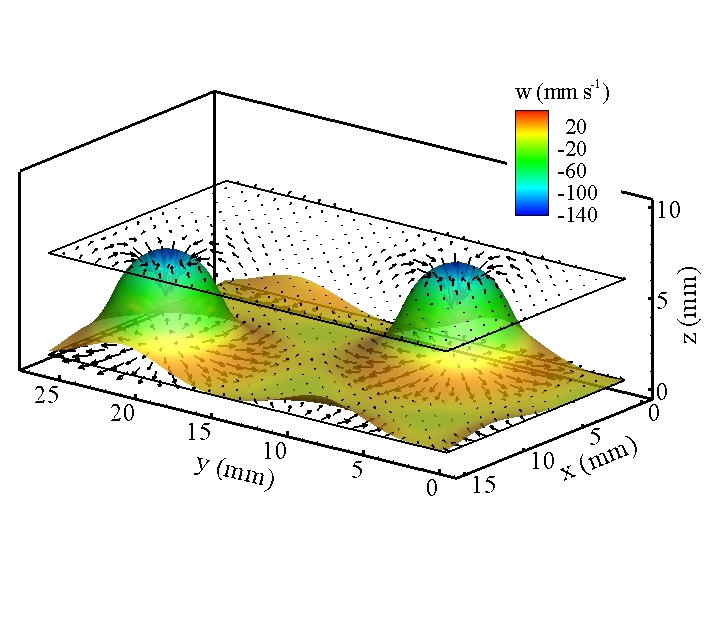}
\vspace*{-2cm}
\caption{Velocity field at time $t=0.07 \times (2T)$ after 
the instant of maximum height. Interface is colored according 
to the vertical velocity $w$. Arrows show velocity field at 
$z=0.53\mm$ and $z=6.08\mm$.
(Total height is $10\mm$, average interface height is $1.6\mm$.)
For clarity, 
velocity vectors are plotted only at every fourth gridpoint in each direction.
Note that the vertical and horizontal scales are different.
One computational domain is shown.}
\label{fig:vel07}
\includegraphics[width=12cm]{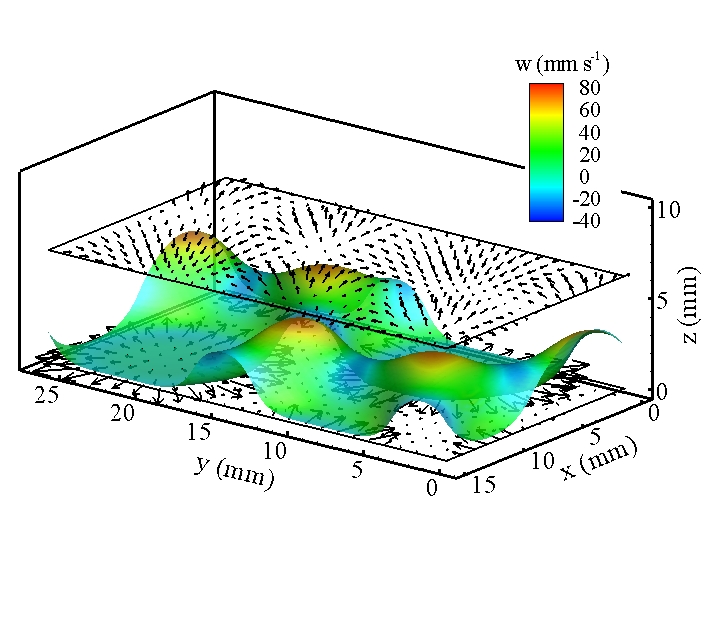}
\vspace*{-2cm}
\caption{Velocity field at time $t=0.41 \times (2T)$ after
the instant of maximum height. 
Vectors shown at $z=0.083\mm$ and $z=6.25\mm$.
Vector and color scales differ from those of figure \ref{fig:vel07}.}
\label{fig:vel41}
\end{figure}
\begin{figure}
\includegraphics[width=12cm]{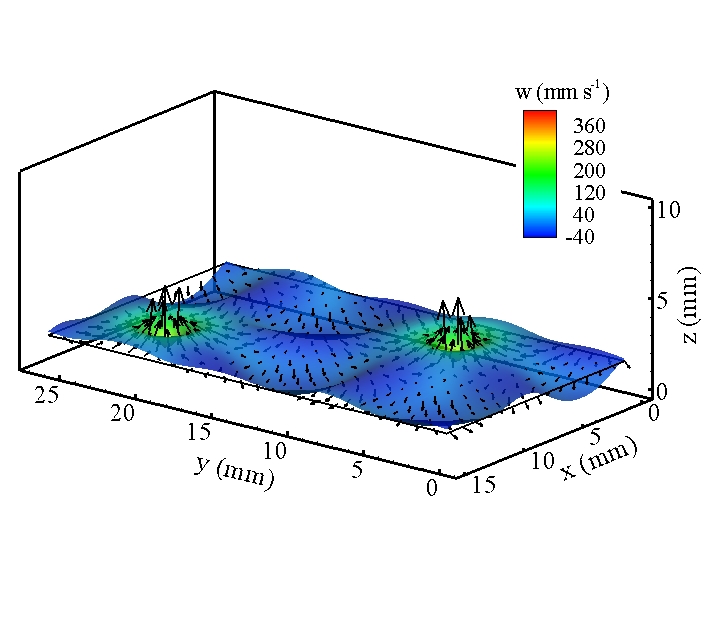}
\vspace*{-2cm}
\caption{Velocity field at time $t=0.73 \times (2T)$ after
the instant of maximum height. Arrows show velocity field at 
$z=1.58\mm$.
Vector and color scales differ from those of figures \ref{fig:vel07}
and \ref{fig:vel41}.}
\label{fig:vel73}
\end{figure}

In addition to the interface height,
our simulations also produce the entire velocity field,
which is the focus of figures \ref{fig:vel07}--\ref{fig:vel73}.
These figures show the velocity fields 
on horizontal planes at three instants spanning the oscillation 
period of a hexagonal pattern, as well as the vertical 
velocity on the interface.
Figures \ref{fig:vel07}, \ref{fig:vel41} and \ref{fig:vel73}
correspond approximately to the visualizations of figures \ref{fig9},
\ref{fig10} and \ref{fig10b}, where the structures are more visible 
since the interface has been repeated periodically in the
horizontal directions for clarity.
The parameters are the same as those given previously,
except that the acceleration $a$ has been decreased to
$36.0\mperssq$, and the number of triangles used to represent 
the interface has been increased to 64 times 
the total number of horizontal gridpoints.

Figure \ref{fig:vel07} is taken at $t=0.07 \times (2T)$, just after 
the interface reaches its maximum height (at $t=0$), when 
the peaks are beginning to descend.
Consequently, the fluid converges horizontally towards the interface peaks,
then descends dramatically below them.
The fluid then diverges horizontally outwards near the bottom
and moves upwards in the large regions between the peaks.
The motion shown in figure \ref{fig:vel41}, at $t=0.41\times (2T)$,
is quite different from that in figure \ref{fig:vel07}.
The peaks of figure \ref{fig:vel07} have collapsed into
wide flat craters.
The fluid converges inwards horizontally above the peaks,
then descends into the craters and diverges outwards horizontally just below them.
Figure \ref{fig:vel73}, at $t=0.73\times (2T)$,
shows that the rims of the wide flat craters seen in figure \ref{fig:vel41}
have in turn collapsed inwards, forming circular waves which invade
the craters, whose remnants are visible as dimples.
The velocity field of figure \ref{fig:vel73} shows fluid converging 
horizontally below these dimples.
These are erupting at velocities which are the largest in the cycle,
and will eventually reconstitute
the high peaks seen in figure \ref{fig:vel07}.

Figures \ref{fig:vel07}--\ref{fig:vel73}, as well as figures \ref{fig9}--\ref{fig10b},
show that these cases pose great computational difficulties.
The interface periodically forms a very thin film (approximately $0.1\mm$;
see wide crater in figures \ref{fig10a} and \ref{fig:vel41})
over large portions of the lower boundary,
within which the velocity may be significant.
These features make it difficult to 
adequately resolve the flow in this layer.
For the time being we use a uniform grid spacing; however, in this case 
an adaptive grid would be more efficient and is under development.
We have also simulated hexagons with resolutions of $70\times120\times100$ and
$70\times120\times50$. Although we do not show these, the two highest resolutions lead to very similar spatial spectra with
a maximum difference in amplitudes of the principal modes of about 5\% between the $70\times120\times100$ and 
$58\times100\times180$ resolutions.  The case resolved by only 50 cells in the $z$ direction shows differences mainly in the $2k_c$ mode where the difference between the $70\times120\times50$ and $58\times100\times180$ resolution
is about 25\%; for the two other modes the difference is about 10\%.   Hexagonal motifs were observed for all of the resolutions. 

The calculation for the hexagon case, for the resolution of
$58\times100\times180$ takes about 7 h per subharmonic oscillation on a
2.16 GHz Intel processor. This corresponds to 42 h of calculation time for
1 s of physical time.

In contrast to the square patterns, all of the hexagonal patterns that we have observed are transient. 
In our calculations, they last for several seconds, i.e. about 
15--20 subharmonic oscillation periods, over which time the amplitudes 
and periods of the principal modes remain constant.
This is also the case for the experimental observations (A. Kityk
\& C. Wagner, private communication), although the experimental lifetimes 
are longer. In our simulations, hexagonal patterns alternated with 
patterns with other symmetries, whose lifetimes were long 
(on the order of several seconds) but irregular.
This behaviour suggests that the hexagonal 
state may belong to a heteroclinic orbit. 
A more extensive examination of the hexagonal regime
will be the subject of a future investigation.
 
\section{Conclusion}

We have carried out full nonlinear three-dimensional simulations of Faraday
waves.  The incompressible Navier--Stokes equations for two fluid layers of
different densities and viscosities are solved using a finite-difference
method. The interface motion and surface tension are treated using a
front-tracking/immersed-boundary technique.  The simulations are validated in
several ways. First, for small oscillation amplitudes, our computations match
the solution of \cite{KT1994} to the Floquet problem which results from the
linearized evolution equations.  The boundaries of the instability tongues,
i.e. the critical amplitude as a function of horizontal wavenumber are
calculated for several wavenumbers on several tongues and are in good
agreement with the theoretical values.  The temporal dependence of the Floquet
modes is also well reproduced by our numerical results, an even more
quantitatively significant validation.

For finite oscillation amplitude, our computations reproduce the square and
hexagonal patterns observed by Kityk \etal~(2005, 2009) at moderate and high-oscillation amplitudes, respectively.  Although the domains shown in figure
\ref{fig8} were chosen to accommodate square and hexagonal patterns
respectively, we consider the emergence of these patterns at the appropriate
parameter values a non-trivial test of our program, since these domains can
also accommodate rectangles and stripes.  Quantitative comparisons were made
between experiment and simulation of the spatio-temporal spectra. Our
numerical results lie well within the experimental uncertainty.  The hexagonal
patterns are long-lived transients and show intriguing dynamical behavior.
Our direct numerical simulations provide velocity fields and pressure
throughout the entire domain of calculation.  Thus, we have been able to
ascertain precisely the fluid motion for the Faraday waves, both above and
below the interface between the two fluids.

Our future studies of Faraday waves will include a more
detailed investigation of the dynamics of the hexagonal patterns, and
the simulation and interpretation of oscillons.

\section*{Acknowledgments}

The authors acknowledge J. Fineberg, E. Knobloch and A. Rucklidge for insights
on theoretical aspects of the Faraday phenomenon, J. Chergui, M. Firdaouss and
K. Boro\'nska for advice regarding implementation of certain numerical
algorithms and, especially, A. Kityk and C. Wagner for extensive discussions
and for sharing their experimental data. 

\end{document}